\documentclass{ieeeaccess}
\usepackage{cite}
\usepackage{amsmath,amssymb,amsfonts}
\usepackage{algorithmic}
\usepackage{graphicx}
\usepackage{textcomp}
\usepackage{color}

\usepackage{times}
\usepackage{soul}
\usepackage{url}
\usepackage[utf8]{inputenc}
\usepackage[small]{caption}
\usepackage{booktabs}
\usepackage{algorithm}

\usepackage{calc}

\usepackage{multirow}
\usepackage{rotating}

\makeatletter
\def\Hline{%
\noalign{\ifnum0=`}\fi\hrule \@height 1pt \futurelet
\reserved@a\@xhline}
\makeatother

\def\BibTeX{{\rm B\kern-.05em{\sc i\kern-.025em b}\kern-.08em
    T\kern-.1667em\lower.7ex\hbox{E}\kern-.125emX}}
    
\newcommand{\blockb}[1]{\multirow{3}{*}{ResBlock\(\left[\begin{array}{c}\text{1$\times$1}\\[-.1em] \text{3$\times$3}\\[-.1em] \text{1$\times$1}\end{array}\right]\)$\times$#1}
}

\newcommand{\blockc}[1]{\multirow{2}{*}{ResBlock\(\left[\begin{array}{c}\text{3$\times$3}\\[-.2em] \text{3$\times$3}\\[-.2em]\end{array}\right]\)$\times$#1}
}

\begin{document}

\title{Combining Noise-to-Image and Image-to-Image GANs: Brain MR Image Augmentation for Tumor Detection}
\author{\uppercase{CHANGHEE HAN}\authorrefmark{1,2,3}, \uppercase{LEONARDO RUNDO}\authorrefmark{3,4,5}, \uppercase{RYOSUKE ARAKI}\authorrefmark{6}, \uppercase{YUDAI NAGANO}\authorrefmark{1},\\ \uppercase{YUJIRO FURUKAWA}\authorrefmark{7}, \uppercase{GIANCARLO MAURI}\authorrefmark{5}, \uppercase{HIDEKI NAKAYAMA}\authorrefmark{1\textcolor{black}{,8}}, \uppercase{HIDEAKI HAYASHI}\authorrefmark{2,\textcolor{black}{9}}}
\address[1]{Machine Perception Group, Graduate School of Information Science and Technology, The University of Tokyo, Tokyo 113-8657, Japan}
\address[2]{Research Center for Medical Big Data, National Institute of Informatics, Tokyo 100-0003, Japan}
\address[3]{Department of Radiology, University of Cambridge, Cambridge CB2 0QQ, United Kingdom}
\address[4]{Cancer Research UK Cambridge Institute, Cambridge CB2 0RE, United Kingdom}
\address[5]{Department of Informatics, Systems and Communication, University of Milano-Bicocca, Milan 20126, Italy}
\address[6]{Machine Perception and Robotics Group, Graduate School of Engineering, Chubu University, Aichi 487-8501, Japan}
\address[7]{Department of Psychiatry, Jikei University School of Medicine, Tokyo 105-8461, Japan}
\address[\textcolor{black}{8}]{\textcolor{black}{International Research Center for Neurointelligence (WPI-IRCN), Institutes for Advanced Study, The University of Tokyo, Tokyo 113-8657, Japan}}
\address[\textcolor{black}{9}]{Human Interface Laboratory, Department of Advanced Information Technology, Kyushu University, Fukuoka 819-0395, Japan}

\markboth
{C. Han \headeretal: Combining Noise-to-Image and Image-to-Image GANs}
{C. Han \headeretal: Combining Noise-to-Image and Image-to-Image GANs}

\corresp{Corresponding author: Changhee Han (e-mail: han@nlab.ci.i.u-tokyo.ac.jp).\vspace{-0.10in}}

\begin{abstract}
Convolutional Neural Networks (CNNs) achieve excellent computer-assisted diagnosis \textcolor{black}{with} sufficient annotated training data. \textcolor{black}{However}, most medical imaging datasets are small and fragmented. In this context, Generative Adversarial Networks (GANs) can synthesize realistic/diverse additional training images to fill the data lack in the real image distribution; researchers have improved classification by augmenting \textcolor{black}{data} with noise-to-image (e.g., random noise samples to diverse pathological images) or image-to-image GANs (e.g., a benign image to a malignant one). Yet, no research has reported results combining noise-to-image and image-to-image GANs for further performance boost. Therefore, to maximize the DA effect with the GAN combinations, we propose a two-step GAN-based DA that generates and refines brain \textcolor{black}{Magnetic Resonance (MR)} images with/without tumors separately: (\textit{i}) Progressive Growing of GANs (PGGANs), multi-stage noise-to-image GAN for high-resolution \textcolor{black}{MR} image generation, first generates realistic/diverse $256\times256$ images; (\textit{ii}) \textcolor{black}{Multimodal UNsupervised Image-to-image Translation (MUNIT) that combines GANs/Variational AutoEncoders or SimGAN that uses a DA-focused GAN loss}, further refines the texture/shape of the PGGAN-generated images similarly to the real ones. We thoroughly investigate CNN-based tumor classification results, also considering the influence of pre-training on ImageNet and discarding weird-looking GAN-generated images. The results show that, when combined with classic DA, our two-step GAN-based DA can significantly outperform the classic DA alone, in tumor detection (i.e., boosting sensitivity \textcolor{black}{$93.67\%$ to $97.48\%$}) and also in other medical imaging tasks.
\end{abstract}



\titlepgskip=-15pt

\maketitle

\section{Introduction}
\label{sec:intro}
Convolutional Neural Networks (CNNs) are playing a key role in medical image analysis, updating the state-of-the-art in many tasks~\cite{Havaei, Rundo, Ker} when large-scale annotated training data are available. However, preparing such massive medical data is demanding; thus, for better diagnosis, researchers generally adopt classic Data Augmentation (DA) techniques, such as geometric/intensity transformations of original images~\cite{Ronneberger,Milletari}. Those augmented images, however, intrinsically have a similar distribution to the original ones, resulting in limited performance improvement. In this sense, Generative Adversarial Network (GAN)-based DA can considerably increase the performance~\cite{Goodfellow}; since the generated images are realistic but completely new samples, they can fill the real image distribution uncovered by the original dataset~\textcolor{black}{\cite{Yi}}.

The main problem in computer-assisted diagnosis lies in small/fragmented medical imaging datasets from \textcolor{black}{multiple} scanners; thus, researchers have improved classification by augmenting images with noise-to-image GANs (e.g., random noise samples to diverse pathological images~\cite{Han1}) or image-to-image GANs (e.g., a benign image to a malignant one~\cite{Wu}). However, no research has \textcolor{black}{achieved further performance boost} by combining noise-to-image and image-to-image GANs.

\begin{figure}[t]
  \centering
  \centerline{\includegraphics[width=\columnwidth]{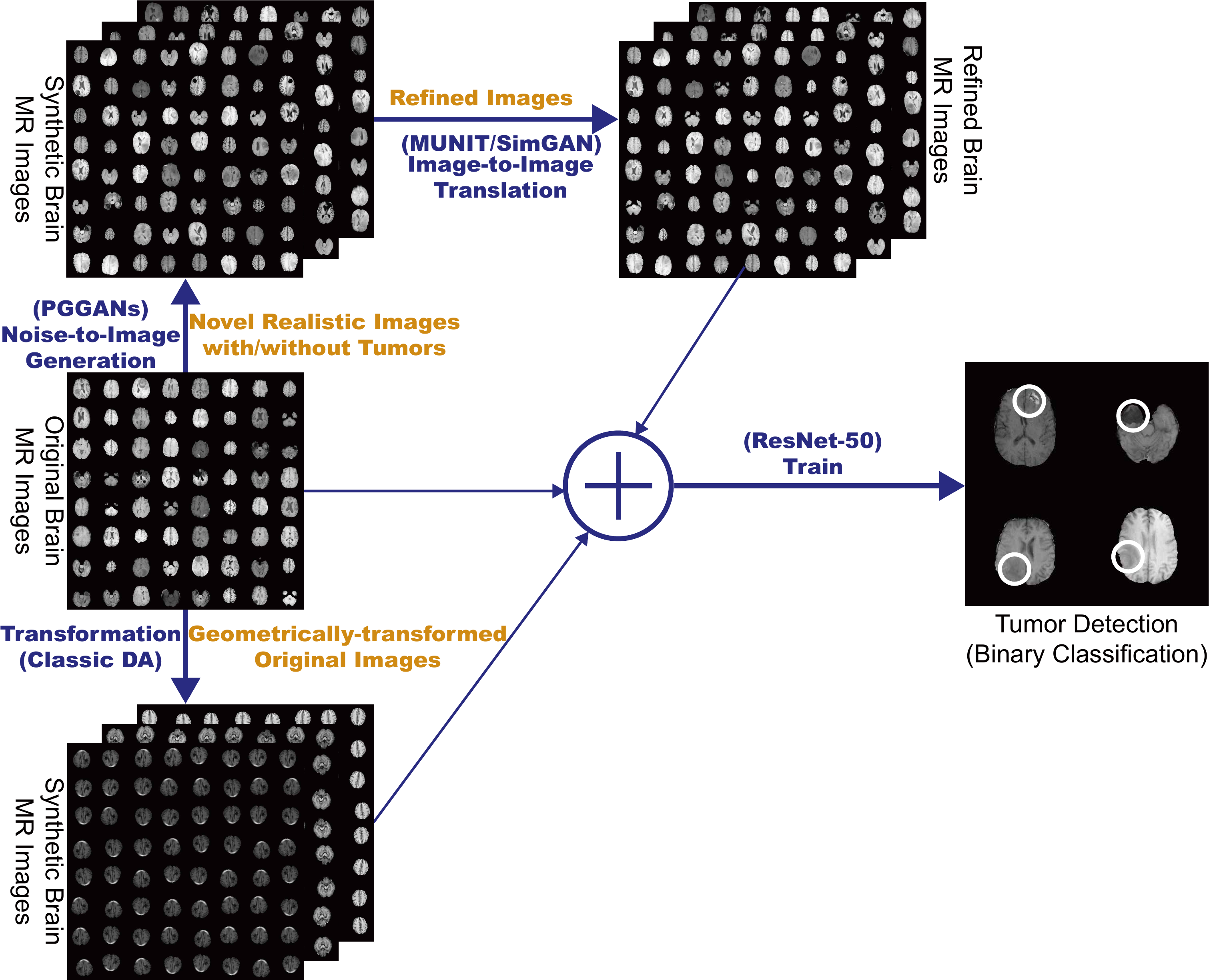}}
\caption{Combining noise-to-image and image-to-image GAN for better tumor detection: the PGGANs generates a number
of realistic brain tumor/non-tumor MR images separately, the \textcolor{black}{MUNIT}/SimGAN refines them separately, and the binary classifier uses them as additional training data.}
\label{fig1}
\vspace{-0.05in}
\end{figure}

So, how can we maximize the DA effect under limited training images using the GAN combinations? To generate and refine brain \textcolor{black}{Magnetic Resonance (MR)} images with/without tumors separately, we propose a two-step GAN-based DA approach: (\textit{i}) Progressive Growing of GANs (PGGANs)~\cite{Karras}, low-to-high resolution noise-to-image GAN, first generates realistic/diverse $256\times256$ images---the PGGANs \textcolor{black}{helps DA} since most CNN architectures adopt around $256\times256$ input sizes (e.g., InceptionResNetV2~\cite{Szegedy}: $299\times299$, ResNet-50~\cite{He}: $224\times224$); (\textit{ii}) \textcolor{black}{Multimodal UNsupervised Image-to-image Translation (MUNIT)~\cite{Huang} that combines GANs/Variational AutoEncoders (VAEs)~\cite{Kingma} or SimGAN~\cite{Shrivastava} that uses a DA-focused GAN loss}, further refines the texture/shape of the PGGAN-generated images to fit them into the real image distribution. \textcolor{black}{Since training a single sophisticated GAN system is already difficult, instead of end-to-end training, we adopt a two-step approach for performance boost \textit{via} an ensemble generation process from those state-of-the-art GANs' different algorithms.}

We thoroughly investigate CNN-based tumor classification results, also considering the influence of pre-training on ImageNet~\cite{Russakovsky} and discarding weird-looking GAN-generated images. Moreover, we evaluate the synthetic images' \textcolor{black}{appearance} \textit{via} Visual Turing Test~\cite{Salimans} by an expert physician, and visualize the data distribution of real/synthetic images \textit{via} t-Distributed Stochastic Neighbor Embedding (t-SNE)~\cite{van der Maaten}. When combined with classic DA, our two-step GAN-based DA approach significantly outperforms the classic DA alone, boosting sensitivity \textcolor{black}{$93.67\%$ to $97.48\%$}\footnote{This paper remarkably improves our preliminary work~\cite{Han1} investigating the potential of the \textcolor{black}{ImageNet-pre-trained} PGGANs---with minimal pre-processing and no refinement---for DA using a vanilla version of ResNet-50\textcolor{black}{, resulting in minimum performance boost; since PGGAN-generated images unstabilized ResNet-50 training, we further optimize the ResNet-50 hyper-parameters (i.e., the optimizer, learning rate, and decay rate) according to the training data, also modifying its architecture before the final sigmoid layer.}}.

\vspace{0.05in}

\noindent \textbf{Research Questions.} We mainly address two questions:
\begin{itemize}
\item \textbf{GAN Selection:} Which GAN architectures are well-suited for realistic/diverse medical image generation?

\item \textbf{Medical DA:} How to use GAN-generated images as additional training data for better CNN-based diagnosis?

\end{itemize}

\noindent \textbf{Contributions.} Our main contributions are as follows:
\begin{itemize}
\item \textbf{Whole Image Generation:} This research shows that PGGANs can generate realistic/diverse $256 \times 256$ whole medical images\textcolor{black}{---not only small pathological sub-areas---and \textcolor{black}{MUNIT} can further refine their texture/shape similarly to real ones}.

\item \textbf{Two-step GAN-based DA:} This novel two-step approach, combining for the first time noise-to-image and image-to-image GANs, \textcolor{black}{significantly} boosts tumor detection \textcolor{black}{sensitivity}.

\item \textbf{Misdiagnosis Prevention:} This study firstly analyzes how medical GAN-based DA is associated with pre-training on ImageNet and discarding weird-looking synthetic images to achieve high sensitivity with small/fragmented datasets.

\end{itemize}

\textcolor{black}{The manuscript is organized as follows.
Section~\ref{sec:GANs} covers the background of GANs, especially focusing on GAN-based DA in medical imaging. Section~\ref{sec:MatMeth} describes the analyzed brain tumor MRI dataset, along with the investigated image generation method using a noise-to-image GAN (i.e., PGGANs) and refinement methods using image-to-image GANs (i.e., \textcolor{black}{MUNIT} and SimGAN), respectively. This section also explains how to evaluate those synthesized images based on tumor detection \textit{via} ResNet-50, clinical validation \textit{via} Visual Turing Test, and visualization \textit{via} t-SNE.
Section~\ref{sec:results} presents and discusses the experimental results.
Lastly, Section~\ref{sec:conclusions} provides the conclusive remarks and future directions.}

\section{Generative Adversarial Networks}
\label{sec:GANs}
VAEs~\cite{Kingma} often \textcolor{black}{accompany} blurred samples despite easier training, due to the imperfect reconstruction using a single objective function; meanwhile, GANs~\cite{Goodfellow} have revolutionized image generation in terms of realism/diversity~\cite{Zhu} based on a two-player objective function: a generator $G$ tries to generate realistic images to fool a discriminator $D$ while maintaining diversity; $D$ attempts to distinguish between the real/\textcolor{black}{synthetic images}. However, difficult GAN training from the two-player objective function accompanies artifacts/mode collapse~\cite{Gulrajani}, when generating high-resolution images (e.g., $256 \times 256$ pixels)~\cite{Radford}; to tackle this, multi-stage noise-to-image GANs have been proposed: AttnGAN~\cite{Xu} generates images from text using attention-based multi-stage refinement; PGGANs~\cite{Karras} generates realistic images using \textcolor{black}{low-to-high resolution} multi-stage training. Contrarily, to obtain images with desired texture/shape, researchers have proposed image-to-image GANs: \textcolor{black}{MUNIT}~\cite{Huang} translates images using both GANs/VAEs; SimGAN~\cite{Shrivastava} translates images for DA using the self-regularization term/local adversarial loss.

Especially in medical imaging, to handle small and fragmented datasets from multiple scanners, researchers have exploited both noise-to-image and image-to-image GANs as DA techniques to improve classification: researchers used the noise-to-image GANs to augment liver lesion Computed Tomography (CT)~\cite{Frid-Adar} and chest cardiovascular abnormality X-ray images~\cite{Madani}; others used the image-to-image GANs to augment breast cancer mammography images~\cite{Wu} and bone lesion X-ray images~\cite{Gupta}, translating benign images \textcolor{black}{into} malignant ones and \textit{vice versa}.

However, to the best of our knowledge, we are the first to combine noise-to-image and image-to-image GANs to maximize the DA performance. Moreover, this is the first medical GAN work generating whole $256 \times 256$ images, instead of regions of interest (i.e., small pathological \textcolor{black}{sub-areas}) alone, for robust classification. Along with classic image transformations, a novel approach---augmenting realistic/diverse whole medical images with the two-step GAN---may become a clinical breakthrough.

\section{Materials and Methods}
\label{sec:MatMeth}

\subsection{BRATS 2016 Training Set}
We use a dataset of $240 \times 240$ contrast-enhanced T1-weighted (T1c) brain axial MR images of $220$ high-grade glioma cases from the Multimodal Brain Tumor Image Segmentation Benchmark (BRATS) 2016~\cite{Menze}. T1c is the most common sequence in tumor detection thanks to its high-contrast~\cite{Koley}.

\begin{figure}[t]
  \centering
  \centerline{\includegraphics[width=\columnwidth]{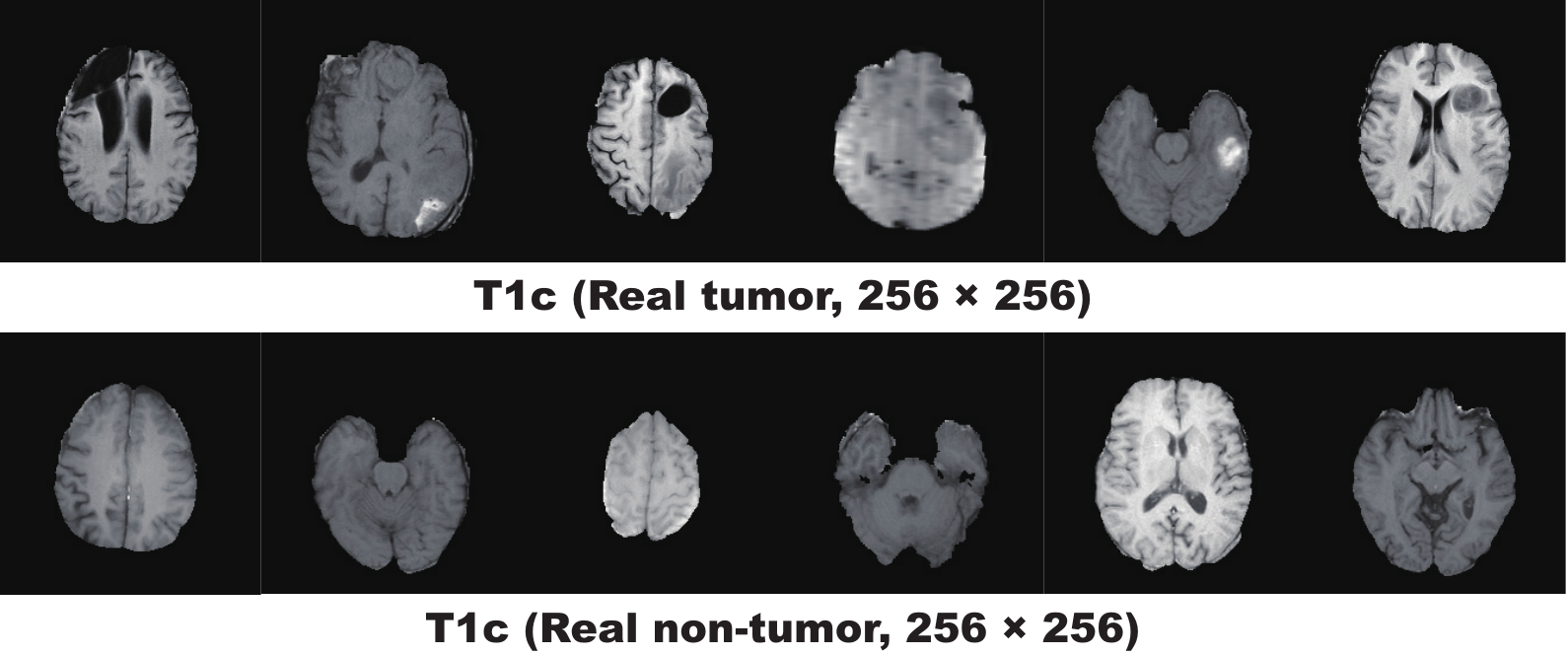}}
\caption{Example real MR images used for PGGAN training.}
\vspace{-0.1in}
\label{fig2}
\end{figure}
\vspace{-0.05in}
\subsection{PGGAN-based Image Generation}
\noindent \textbf{Pre-processing}
For better GAN/ResNet-50 training, we select the slices from $\#30$ to $\#130$ among the whole $155$ slices to omit initial/final slices, which convey negligible useful information; also, since tumor/non-tumor annotation in the BRATS 2016 dataset, based on 3D volumes, is highly incorrect/ambiguous on 2D slices, we exclude ($i$) tumor images tagged as non-tumor, ($ii$) non-tumor images tagged as tumor, ($iii$) borderline images with unclear tumor/non-tumor appearance, and ($iv$) images with missing brain parts due to the skull-stripping procedure\footnote{Although this discarding procedure could be automated, we manually conduct it for reliability.}. For tumor detection, we divide the whole dataset ($220$ patients) into:


\begin{itemize}
\item Training set \\($154$ patients/$4,679$ tumor/$3,750$ non-tumor images);
\item Validation set \\($44$ patients/$750$ tumor/$608$ non-tumor images);
\item Test set \\($22$ patients/$1,232$ tumor/$1,013$ non-tumor images).
\end{itemize}

During the GAN training, we only use the training set to be fair; for better \textcolor{black}{PGGAN} training, the training set images are zero-padded to reach a power of $2$: $256 \times 256$ pixels from $240 \times 240$. Fig.~\ref{fig2} shows example real MR images.

\begin{figure}[t]
  \centering
  \centerline{\includegraphics[width=\columnwidth]{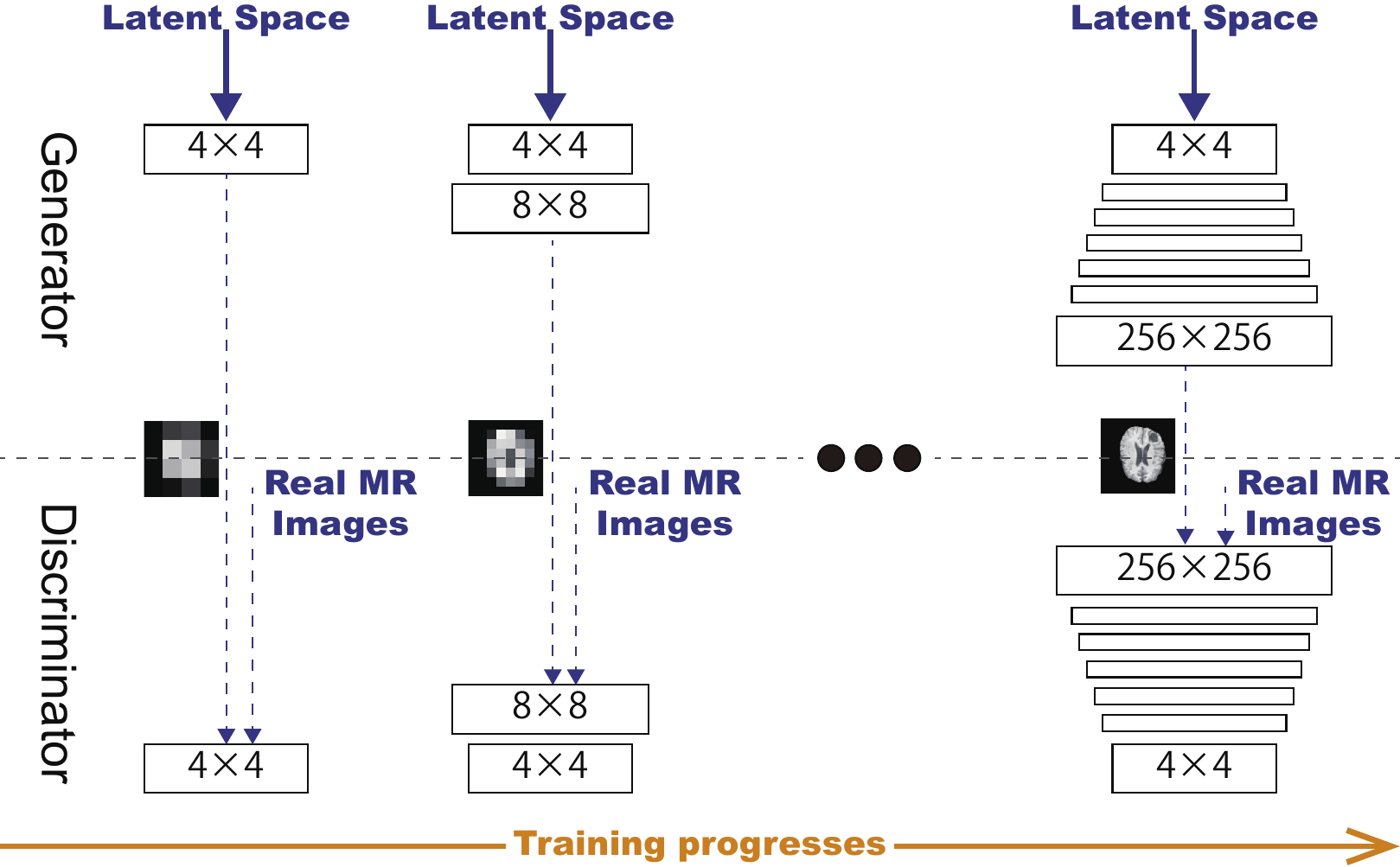}}
\caption{PGGAN architecture for $256 \times 256$ brain MR image generation. \textcolor{black}{$N \times N$ refers to convolutional layers operating on $N \times N$ spatial resolution.}}
\label{fig3}
\end{figure}

\vspace{0.1in}
\vspace{-0.05in}
\noindent \textbf{PGGANs}~\cite{Karras} is a GAN training method that progressively grows a generator and discriminator: starting from low resolution, new layers model details as training progresses. This study adopts the PGGANs to synthesize realistic/diverse $256 \times 256$ brain MR images (Fig.~\ref{fig3}); we train and generate tumor/non-tumor images separately.

\vspace{0.1in}
\vspace{-0.05in}
\noindent \textbf{PGGAN Implementation Details} The PGGAN architecture adopts the Wasserstein loss with gradient penalty~\cite{Gulrajani}:
\begin{eqnarray}\label{eq:wgan_gp}
\underset{{\tilde{\mathbf{y}}\sim{\mathbb{P}_g}}}{\mathbb{E}}[D(\tilde{\mathbf{y}})]-\underset{{\mathbf{y}\sim{\mathbb{P}_r}}}{\mathbb{E}}[D(\mathbf{y})] +
\lambda_\text{gp}\underset{{{\hat{\mathbf{y}}}\sim{\mathbb{P}_{\hat{\mathbf{y}}}}}} {\mathbb{E}}[(\left \| \nabla_{\hat{\mathbf{y}}}{D({\hat{\mathbf{y}}})} \right \|_2-1)^2],
\end{eqnarray}
\textcolor{black}{where $\mathbb{E}[\cdot]$ denotes the expected value, the discriminator $D \in \mathcal{D}$ (i.e., the set of $1$-Lipschitz functions)}, $\mathbb{P}_r$ is the data distribution defined by the true data sample $\mathbf{y}$, and $\mathbb{P}_g$ is the model distribution defined by the generated sample \textcolor{black}{${\tilde{\mathbf{y}} = G(\mathbf{z})}$ ($\mathbf{z} \sim p(\mathbf{z})$ is the input noise $\mathbf{z}$ to the generator sampled from a Gaussian distribution)}. A gradient penalty is added for the random sample ${\hat{\mathbf{y}}}\sim{\mathbb{P}_{\hat{\mathbf{y}}}}$, \textcolor{black}{where $\nabla_{\hat{\mathbf{y}}}$ is the gradient operator towards the generated samples} and $\lambda_\text{gp}$ is the gradient penalty coefficient.


We train the model \textcolor{black}{(Table~\ref{tab1})} for $100$ epochs with a batch size of $16$ and $1.0 \times 10^{-3}$ learning rate for the Adam optimizer \textcolor{black}{(the exponential decay rates $\beta_{1} = 0, \beta_{2} = 0.99$)}~\cite{Kingma2}. \textcolor{black}{All experiments use $\lambda_\text{gp} = 10$ with $1$ critic iteration per generator iteration.} During training, we apply random cropping in $0$-$15$ pixels \textcolor{black}{as DA.}

\begin{table*}[!t]
\caption{\textcolor{black}{PGGAN architecture details for the generator/discriminator. Pixelwise feature vector normalization\cite{Krizhevsky} is applied in the generator after each convolutional layer except for the final output layer as in the original paper~\cite{Karras}. LReLU denotes Leaky ReLU with leakiness $0.2$.}}
\small
  \centering
\begin{tabular}{lcc}
\Hline
\textbf{\textcolor{black}{Generator}} & \textbf{\textcolor{black}{Activation}} & \textbf{\textcolor{black}{Output Shape}}\\
\hline
\textcolor{black}{Latent vector} & \textcolor{black}{--} & \textcolor{black}{$\makebox[\widthof{512}][c]{512}\times\makebox[\widthof{1024}][c]{1}\times\makebox[\widthof{1024}][c]{1}$} \\
\textcolor{black}{Conv $4\times4$} & \textcolor{black}{LReLU} & \textcolor{black}{$\makebox[\widthof{512}][c]{512}\times\makebox[\widthof{1024}][c]{4}\times\makebox[\widthof{1024}][c]{4}$} \\
\textcolor{black}{Conv $3\times3$} &  \textcolor{black}{LReLU} & \textcolor{black}{$\makebox[\widthof{512}][c]{512}\times\makebox[\widthof{1024}][c]{4}\times\makebox[\widthof{1024}][c]{4}$} \\
\hline
\textcolor{black}{Upsample} & \textcolor{black}{--} & \textcolor{black}{$\makebox[\widthof{512}][c]{512}\times\makebox[\widthof{1024}][c]{8}\times\makebox[\widthof{1024}][c]{8}$} \\
\textcolor{black}{Conv $3\times3$} & \textcolor{black}{LReLU} & \textcolor{black}{$\makebox[\widthof{512}][c]{512}\times\makebox[\widthof{1024}][c]{8}\times\makebox[\widthof{1024}][c]{8}$} \\
\textcolor{black}{Conv $3\times3$} & \textcolor{black}{LReLU} & \textcolor{black}{$\makebox[\widthof{512}][c]{512}\times\makebox[\widthof{1024}][c]{8}\times\makebox[\widthof{1024}][c]{8}$} \\
\hline
\textcolor{black}{Upsample} & \textcolor{black}{--} & \textcolor{black}{$\makebox[\widthof{512}][c]{512}\times\makebox[\widthof{1024}][c]{16}\times\makebox[\widthof{1024}][c]{16}$}  \\
\textcolor{black}{Conv $3\times3$} & \textcolor{black}{LReLU} & \textcolor{black}{$\makebox[\widthof{512}][c]{256}\times\makebox[\widthof{1024}][c]{16}\times\makebox[\widthof{1024}][c]{16}$} \\
\textcolor{black}{Conv $3\times3$} & \textcolor{black}{LReLU} & \textcolor{black}{$\makebox[\widthof{512}][c]{256}\times\makebox[\widthof{1024}][c]{16}\times\makebox[\widthof{1024}][c]{16}$} \\
\hline
\textcolor{black}{Upsample} & \textcolor{black}{--} & \textcolor{black}{$\makebox[\widthof{512}][c]{256}\times\makebox[\widthof{1024}][c]{32}\times\makebox[\widthof{1024}][c]{32}$}  \\
\textcolor{black}{Conv $3\times3$} & \textcolor{black}{LReLU} & \textcolor{black}{$\makebox[\widthof{512}][c]{128}\times\makebox[\widthof{1024}][c]{32}\times\makebox[\widthof{1024}][c]{32}$} \\
\textcolor{black}{Conv $3\times3$} & \textcolor{black}{LReLU} & \textcolor{black}{$\makebox[\widthof{512}][c]{128}\times\makebox[\widthof{1024}][c]{32}\times\makebox[\widthof{1024}][c]{32}$} \\
\hline
\textcolor{black}{Upsample} & \textcolor{black}{--} & \textcolor{black}{$\makebox[\widthof{512}][c]{128}\times\makebox[\widthof{1024}][c]{64}\times\makebox[\widthof{1024}][c]{64}$} \\ 
\textcolor{black}{Conv $3\times3$} & \textcolor{black}{LReLU} & \textcolor{black}{$\makebox[\widthof{512}][c]{64}\times\makebox[\widthof{1024}][c]{64}\times\makebox[\widthof{1024}][c]{64}$} \\
\textcolor{black}{Conv $3\times3$} & \textcolor{black}{LReLU} & \textcolor{black}{$\makebox[\widthof{512}][c]{64}\times\makebox[\widthof{1024}][c]{64}\times\makebox[\widthof{1024}][c]{64}$} \\
\hline
\textcolor{black}{Upsample} & \textcolor{black}{--} & \textcolor{black}{$\makebox[\widthof{512}][c]{64}\times\makebox[\widthof{1024}][c]{128}\times\makebox[\widthof{1024}][c]{128}$} \\
\textcolor{black}{Conv $3\times3$} & \textcolor{black}{LReLU} & \textcolor{black}{$\makebox[\widthof{512}][c]{32}\times\makebox[\widthof{1024}][c]{128}\times\makebox[\widthof{1024}][c]{128}$}  \\
\textcolor{black}{Conv $3\times3$} & \textcolor{black}{LReLU} & \textcolor{black}{$\makebox[\widthof{512}][c]{32}\times\makebox[\widthof{1024}][c]{128}\times\makebox[\widthof{1024}][c]{128}$}  \\
\hline
\textcolor{black}{Upsample} & \textcolor{black}{--} & \textcolor{black}{$\makebox[\widthof{512}][c]{32}\times\makebox[\widthof{1024}][c]{256}\times\makebox[\widthof{1024}][c]{256}$} \\
\textcolor{black}{Conv $3\times3$} & \textcolor{black}{LReLU} & \textcolor{black}{$\makebox[\widthof{512}][c]{16}\times\makebox[\widthof{1024}][c]{256}\times\makebox[\widthof{1024}][c]{256}$}  \\
\textcolor{black}{Conv $3\times3$} & \textcolor{black}{LReLU} & \textcolor{black}{$\makebox[\widthof{512}][c]{16}\times\makebox[\widthof{1024}][c]{256}\times\makebox[\widthof{1024}][c]{256}$}  \\
\hline
\textcolor{black}{Conv $1\times1$} & \textcolor{black}{Linear} & \textcolor{black}{$\makebox[\widthof{512}][c]{1}\times\makebox[\widthof{1024}][c]{256}\times\makebox[\widthof{1024}][c]{256}$}  \\
\Hline
\end{tabular}
\hspace{1cm}
\begin{tabular}{lcc}
\Hline
\textcolor{black}{\textbf{Discriminator}} & \textcolor{black}{\textbf{Activation}} & \textcolor{black}{\textbf{Output Shape}}\\
\hline
\textcolor{black}{Input image} & \textcolor{black}{--} & \textcolor{black}{$\makebox[\widthof{512}][c]{1}\times\makebox[\widthof{1024}][c]{256}\times\makebox[\widthof{1024}][c]{256}$} \\
\textcolor{black}{Conv $1\times1$} & \textcolor{black}{LReLU} & \textcolor{black}{$\makebox[\widthof{512}][c]{16}\times\makebox[\widthof{1024}][c]{256}\times\makebox[\widthof{1024}][c]{256}$} \\
\textcolor{black}{Conv $3\times3$} &  \textcolor{black}{LReLU} & \textcolor{black}{$\makebox[\widthof{512}][c]{16}\times\makebox[\widthof{1024}][c]{256}\times\makebox[\widthof{1024}][c]{256}$} \\
\textcolor{black}{Conv $3\times3$} &  \textcolor{black}{LReLU} & \textcolor{black}{$\makebox[\widthof{512}][c]{32}\times\makebox[\widthof{1024}][c]{256}\times\makebox[\widthof{1024}][c]{256}$} \\
\textcolor{black}{Downsample} & \textcolor{black}{--} & \textcolor{black}{$\makebox[\widthof{512}][c]{32}\times\makebox[\widthof{1024}][c]{128}\times\makebox[\widthof{1024}][c]{128}$} \\
\hline
\textcolor{black}{Conv $3\times3$} & \textcolor{black}{LReLU} & \textcolor{black}{$\makebox[\widthof{512}][c]{32}\times\makebox[\widthof{1024}][c]{128}\times\makebox[\widthof{1024}][c]{128}$} \\
\textcolor{black}{Conv $3\times3$} & \textcolor{black}{LReLU}& \textcolor{black}{$\makebox[\widthof{512}][c]{64}\times\makebox[\widthof{1024}][c]{128}\times\makebox[\widthof{1024}][c]{128}$} \\
\textcolor{black}{Downsample} & \textcolor{black}{--} & \textcolor{black}{$\makebox[\widthof{512}][c]{64}\times\makebox[\widthof{1024}][c]{64}\times\makebox[\widthof{1024}][c]{64}$}  \\
\hline
\textcolor{black}{Conv $3\times3$} & \textcolor{black}{LReLU} & \textcolor{black}{$\makebox[\widthof{512}][c]{64}\times\makebox[\widthof{1024}][c]{64}\times\makebox[\widthof{1024}][c]{64}$} \\
\textcolor{black}{Conv $3\times3$} & \textcolor{black}{LReLU} & \textcolor{black}{$\makebox[\widthof{512}][c]{128}\times\makebox[\widthof{1024}][c]{64}\times\makebox[\widthof{1024}][c]{64}$} \\
\textcolor{black}{Downsample} & \textcolor{black}{--} & \textcolor{black}{$\makebox[\widthof{512}][c]{128}\times\makebox[\widthof{1024}][c]{32}\times\makebox[\widthof{1024}][c]{32}$}  \\
\hline
\textcolor{black}{Conv $3\times3$} & \textcolor{black}{LReLU} & \textcolor{black}{$\makebox[\widthof{512}][c]{128}\times\makebox[\widthof{1024}][c]{32}\times\makebox[\widthof{1024}][c]{32}$} \\
\textcolor{black}{Conv $3\times3$} & \textcolor{black}{LReLU} & \textcolor{black}{$\makebox[\widthof{512}][c]{256}\times\makebox[\widthof{1024}][c]{32}\times\makebox[\widthof{1024}][c]{32}$} \\
\textcolor{black}{Downsample} & \textcolor{black}{--} & \textcolor{black}{$\makebox[\widthof{512}][c]{256}\times\makebox[\widthof{1024}][c]{16}\times\makebox[\widthof{1024}][c]{16}$} \\ 
\hline
\textcolor{black}{Conv $3\times3$} & \textcolor{black}{LReLU} & \textcolor{black}{$\makebox[\widthof{512}][c]{256}\times\makebox[\widthof{1024}][c]{16}\times\makebox[\widthof{1024}][c]{16}$} \\
\textcolor{black}{Conv $3\times3$} & \textcolor{black}{LReLU} & \textcolor{black}{$\makebox[\widthof{512}][c]{512}\times\makebox[\widthof{1024}][c]{16}\times\makebox[\widthof{1024}][c]{16}$} \\
\textcolor{black}{Downsample} & \textcolor{black}{--} & \textcolor{black}{$\makebox[\widthof{512}][c]{512}\times\makebox[\widthof{1024}][c]{8}\times\makebox[\widthof{1024}][c]{8}$} \\
\hline
\textcolor{black}{Conv $3\times3$} & \textcolor{black}{LReLU} & \textcolor{black}{$\makebox[\widthof{512}][c]{512}\times\makebox[\widthof{1024}][c]{8}\times\makebox[\widthof{1024}][c]{8}$}  \\
\textcolor{black}{Conv $3\times3$} & \textcolor{black}{LReLU} & \textcolor{black}{$\makebox[\widthof{512}][c]{512}\times\makebox[\widthof{1024}][c]{8}\times\makebox[\widthof{1024}][c]{8}$}  \\
\textcolor{black}{Downsample} & \textcolor{black}{--} & \textcolor{black}{$\makebox[\widthof{512}][c]{512}\times\makebox[\widthof{1024}][c]{4}\times\makebox[\widthof{1024}][c]{4}$} \\
\hline
\textcolor{black}{Minibatch stddev} & \textcolor{black}{--} & \textcolor{black}{$\makebox[\widthof{512}][c]{513}\times\makebox[\widthof{1024}][c]{4}\times\makebox[\widthof{1024}][c]{4}$}  \\
\textcolor{black}{Conv $3\times3$} & \textcolor{black}{LReLU} & \textcolor{black}{$\makebox[\widthof{512}][c]{512}\times\makebox[\widthof{1024}][c]{4}\times\makebox[\widthof{1024}][c]{4}$} \\
\textcolor{black}{Conv $4\times4$} & \textcolor{black}{LReLU} & \textcolor{black}{$\makebox[\widthof{512}][c]{512}\times\makebox[\widthof{1024}][c]{1}\times\makebox[\widthof{1024}][c]{1}$} \\
\textcolor{black}{Fully-connected} & \textcolor{black}{Linear} & \textcolor{black}{$\makebox[\widthof{512}][c]{1}\times\makebox[\widthof{1024}][c]{1}\times\makebox[\widthof{1024}][c]{1}$}  \\
\Hline
\label{tab1}
\end{tabular}
\end{table*}

\subsection{MUNIT/SimGAN-based Image Refinement}
\noindent \textbf{Refinement}
\textcolor{black}{Using resized $224\times224$ images for ResNet-50,} we further refine the texture/shape of PGGAN-generated tumor/non-tumor images separately to fit them into the real image distribution using \textcolor{black}{MUNIT}~\cite{Huang} or SimGAN~\cite{Shrivastava}. SimGAN remarkably improved eye gaze estimation results after refining non-GAN-based synthetic images from the UnityEyes simulator $via$ image-to-image translation; thus, we also expect such performance improvement after refining synthetic images from a noise-to-image GAN (i.e., PGGANs) $via$ an image-to-image GAN (i.e., \textcolor{black}{MUNIT}/SimGAN) with considerably different GAN algorithms.

We randomly select $3,000$ real/$3,000$ PGGAN-generated tumor images for tumor image training, and we perform the same for non-tumor image training. To find suitable refining steps for each architecture, we pick the \textcolor{black}{MUNIT}/SimGAN models with the highest accuracy on tumor detection validation, when pre-trained and combined with classic DA, among $20,000$/$50,000$/$100,000$ steps, respectively.

\vspace{0.1in}

\noindent \textbf{MUNIT}~\cite{Huang} \textcolor{black}{is an image-to-image GAN based on both auto-encoding/translation; it extends UNIT~\cite{Liu} to increase the generated images' realism/diversity \textit{via} a stochastic model representing continuous output distributions.}


\vspace{0.1in}

\noindent \textbf{MUNIT Implementation Details}
The \textcolor{black}{MUNIT} architecture adopts the following loss:
\begin{eqnarray}\label{eq:unit_loss}
\min_{E_1,E_2,G_1,G_2} \max_{D_1,D_2} 
&&\mathcal{L}_{\text{\tiny VAE}_1} +\mathcal{L}_{\text{\tiny GAN}_1} + \mathcal{L}_{\text{\tiny CC}_1} + \mathcal{L}_{\text{\tiny VGG}_1}\nonumber\\
&+&\mathcal{L}_{\text{\tiny VAE}_2} + \mathcal{L}_{\text{\tiny GAN}_2} + \mathcal{L}_{\text{\tiny CC}_2} + \mathcal{L}_{\text{\tiny VGG}_2},
\end{eqnarray}
\textcolor{black}{where $\mathcal{L}(\cdot)$ denotes the loss function.}
Using the multiple encoders $E_1$/$E_2$, generators $G_1$/$G_2$, discriminators $D_1$/$D_2$, cycle-consistencies CC$_1$/CC$_2$\textcolor{black}{, and domain-invariant perceptions VGG$_1$/VGG$_2$~\cite{Simonyan}}, this framework jointly solves learning problems of the VAE$_1$/VAE$_2$ and GAN$_1$/GAN$_2$ for the image reconstruction streams, image translation streams, cycle-consistency reconstruction streams, \textcolor{black}{and domain-invariant perception streams. Since we do not need the style loss for our experiments, instead of the MUNIT loss, we use the UNIT loss with the perceptual loss for the MUNIT architecture (as in the UNIT authors' GitHub repository).}



%

We train \textcolor{black}{the model (Table~\ref{tab2})} for $100,000$ steps with a batch size of 1 and $1.0 \times 10^{-4}$ learning rate for the Adam optimizer \textcolor{black}{($\beta_{1} = 0.5, \beta_{2} = 0.999$)}~\cite{Kingma2}. The learning rate is reduced by half every $20,000$ steps. \textcolor{black}{We use the following MUNIT weights: the adversarial loss weight $= 1$; the image reconstruction loss weight $ = 10$; the Kullback-Leibler (KL) divergence loss weight for reconstruction $= 0.01$; the cycle consistency loss weight $ = 10$; the KL divergence loss weight for cycle consistency $= 0.01$; the domain-invariant perceptual loss weight $= 1$; \textcolor{black}{the Least Squares GAN objective function for the discriminators~\cite{Mao}}.} During training, we apply horizontal flipping as DA.

\begin{table*}[!t]
\caption{\textcolor{black}{MUNIT architecture details for the generator/discriminator. We input color images (i.e., 3 channels) to use ImageNet initialization. Instance normalization~\cite{Ulyanov}/adaptive instance normalization~\cite{Huang2} are applied in the content encoder/decoder after each convolutional layer respectively except for the final decoder output layer as in the original paper~\cite{Huang}. LReLU denotes Leaky ReLU with leakiness $0.2$.}}
\small
  \centering
\begin{tabular}{lcc}
\Hline
\textbf{\textcolor{black}{Generator}} & \textbf{\textcolor{black}{Activation}} & \textbf{\textcolor{black}{Output Shape}}\\
\textbf{\textcolor{black}{Content Encoder}} &  & \\
\hline
\textcolor{black}{Input image} & \textcolor{black}{--} & \textcolor{black}{$\makebox[\widthof{512}][c]{3}\times\makebox[\widthof{1024}][c]{224}\times\makebox[\widthof{1024}][c]{224}$} \\
\textcolor{black}{Conv $7\times7$} & \textcolor{black}{ReLU} & \textcolor{black}{$\makebox[\widthof{512}][c]{64}\times\makebox[\widthof{1024}][c]{224}\times\makebox[\widthof{1024}][c]{224}$} \\
\textcolor{black}{Conv $4\times4$} & \textcolor{black}{ReLU} & \textcolor{black}{$\makebox[\widthof{512}][c]{128}\times\makebox[\widthof{1024}][c]{112}\times\makebox[\widthof{1024}][c]{112}$} \\
\textcolor{black}{Conv $4\times4$} & \textcolor{black}{ReLU} & \textcolor{black}{$\makebox[\widthof{512}][c]{256}\times\makebox[\widthof{1024}][c]{56}\times\makebox[\widthof{1024}][c]{56}$} \\
\hline
\textcolor{black}{\blockc{4}} & \textcolor{black}{ReLU} & \textcolor{black}{$\makebox[\widthof{512}][c]{256}\times\makebox[\widthof{1024}][c]{56}\times\makebox[\widthof{1024}][c]{56}$} \\
& \textcolor{black}{--} & \textcolor{black}{$\makebox[\widthof{512}][c]{256}\times\makebox[\widthof{1024}][c]{56}\times\makebox[\widthof{1024}][c]{56}$} \\
\hline
\textbf{\textcolor{black}{Decoder}} &  & \\
\hline
\textcolor{black}{\blockc{4}} & \textcolor{black}{ReLU} & \textcolor{black}{$\makebox[\widthof{512}][c]{256}\times\makebox[\widthof{1024}][c]{56}\times\makebox[\widthof{1024}][c]{56}$} \\
& \textcolor{black}{--} & \textcolor{black}{$\makebox[\widthof{512}][c]{256}\times\makebox[\widthof{1024}][c]{56}\times\makebox[\widthof{1024}][c]{56}$} \\
\hline

\textcolor{black}{Upsample} & \textcolor{black}{--} & \textcolor{black}{$\makebox[\widthof{512}][c]{256}\times\makebox[\widthof{1024}][c]{112}\times\makebox[\widthof{1024}][c]{112}$} \\
\textcolor{black}{Conv $5\times5$} & \textcolor{black}{ReLU} & \textcolor{black}{$\makebox[\widthof{512}][c]{128}\times\makebox[\widthof{1024}][c]{112}\times\makebox[\widthof{1024}][c]{112}$} \\
\hline

\textcolor{black}{Upsample} & \textcolor{black}{--} & \textcolor{black}{$\makebox[\widthof{512}][c]{128}\times\makebox[\widthof{1024}][c]{224}\times\makebox[\widthof{1024}][c]{224}$} \\
\textcolor{black}{Conv $5\times5$} & \textcolor{black}{ReLU} & \textcolor{black}{$\makebox[\widthof{512}][c]{64}\times\makebox[\widthof{1024}][c]{224}\times\makebox[\widthof{1024}][c]{224}$} \\
\hline
\textcolor{black}{Conv $7\times7$} & \textcolor{black}{Tanh} & \textcolor{black}{$\makebox[\widthof{512}][c]{3}\times\makebox[\widthof{1024}][c]{224}\times\makebox[\widthof{1024}][c]{224}$}  \\
\Hline
\end{tabular}
\hspace{1cm}
\begin{tabular}{lcc}
\Hline
\textcolor{black}{\textbf{Discriminator}} & \textcolor{black}{\textbf{Activation}} & \textcolor{black}{\textbf{Output Shape}}\\
\hline
\textcolor{black}{Input image} & \textcolor{black}{--} & \textcolor{black}{$\makebox[\widthof{512}][c]{3}\times\makebox[\widthof{1024}][c]{224}\times\makebox[\widthof{1024}][c]{224}$} \\
\textcolor{black}{Conv $4\times4$} & \textcolor{black}{LReLU} & \textcolor{black}{$\makebox[\widthof{512}][c]{64}\times\makebox[\widthof{1024}][c]{112}\times\makebox[\widthof{1024}][c]{112}$} \\
\textcolor{black}{Conv $4\times4$} & \textcolor{black}{LReLU} & \textcolor{black}{$\makebox[\widthof{512}][c]{128}\times\makebox[\widthof{1024}][c]{56}\times\makebox[\widthof{1024}][c]{56}$} \\
\textcolor{black}{Conv $4\times4$} & \textcolor{black}{LReLU} & \textcolor{black}{$\makebox[\widthof{512}][c]{256}\times\makebox[\widthof{1024}][c]{28}\times\makebox[\widthof{1024}][c]{28}$} \\
\textcolor{black}{Conv $4\times4$} & \textcolor{black}{LReLU} & \textcolor{black}{$\makebox[\widthof{512}][c]{512}\times\makebox[\widthof{1024}][c]{14}\times\makebox[\widthof{1024}][c]{14}$} \\
\textcolor{black}{Conv $4\times4$} & \textcolor{black}{--} & \textcolor{black}{$\makebox[\widthof{512}][c]{1}\times\makebox[\widthof{1024}][c]{14}\times\makebox[\widthof{1024}][c]{14}$} \\
\textcolor{black}{AveragePool} & \textcolor{black}{--} & \textcolor{black}{$\makebox[\widthof{512}][c]{3}\times\makebox[\widthof{1024}][c]{112}\times\makebox[\widthof{1024}][c]{112}$} \\
\hline
\textcolor{black}{Conv $4\times4$} & \textcolor{black}{LReLU} & \textcolor{black}{$\makebox[\widthof{512}][c]{64}\times\makebox[\widthof{1024}][c]{56}\times\makebox[\widthof{1024}][c]{56}$} \\
\textcolor{black}{Conv $4\times4$} & \textcolor{black}{LReLU} & \textcolor{black}{$\makebox[\widthof{512}][c]{128}\times\makebox[\widthof{1024}][c]{28}\times\makebox[\widthof{1024}][c]{28}$} \\
\textcolor{black}{Conv $4\times4$} & \textcolor{black}{LReLU} & \textcolor{black}{$\makebox[\widthof{512}][c]{256}\times\makebox[\widthof{1024}][c]{14}\times\makebox[\widthof{1024}][c]{14}$} \\
\textcolor{black}{Conv $4\times4$} & \textcolor{black}{LReLU} & \textcolor{black}{$\makebox[\widthof{512}][c]{512}\times\makebox[\widthof{1024}][c]{7}\times\makebox[\widthof{1024}][c]{7}$} \\
\textcolor{black}{Conv $4\times4$} & \textcolor{black}{--} & \textcolor{black}{$\makebox[\widthof{512}][c]{1}\times\makebox[\widthof{1024}][c]{7}\times\makebox[\widthof{1024}][c]{7}$} \\
\textcolor{black}{AveragePool} & \textcolor{black}{--} & \textcolor{black}{$\makebox[\widthof{512}][c]{3}\times\makebox[\widthof{1024}][c]{56}\times\makebox[\widthof{1024}][c]{56}$} \\
\hline
\textcolor{black}{Conv $4\times4$} & \textcolor{black}{LReLU} & \textcolor{black}{$\makebox[\widthof{512}][c]{64}\times\makebox[\widthof{1024}][c]{28}\times\makebox[\widthof{1024}][c]{28}$} \\
\textcolor{black}{Conv $4\times4$} & \textcolor{black}{LReLU} & \textcolor{black}{$\makebox[\widthof{512}][c]{128}\times\makebox[\widthof{1024}][c]{14}\times\makebox[\widthof{1024}][c]{14}$} \\
\textcolor{black}{Conv $4\times4$} & \textcolor{black}{LReLU} & \textcolor{black}{$\makebox[\widthof{512}][c]{256}\times\makebox[\widthof{1024}][c]{7}\times\makebox[\widthof{1024}][c]{7}$} \\
\textcolor{black}{Conv $4\times4$} & \textcolor{black}{LReLU} & \textcolor{black}{$\makebox[\widthof{512}][c]{512}\times\makebox[\widthof{1024}][c]{3}\times\makebox[\widthof{1024}][c]{3}$} \\
\textcolor{black}{Conv $4\times4$} & \textcolor{black}{--} & \textcolor{black}{$\makebox[\widthof{512}][c]{1}\times\makebox[\widthof{1024}][c]{3}\times\makebox[\widthof{1024}][c]{3}$} \\
\textcolor{black}{AveragePool} & \textcolor{black}{--} & \textcolor{black}{$\makebox[\widthof{512}][c]{3}\times\makebox[\widthof{1024}][c]{28}\times\makebox[\widthof{1024}][c]{28}$} \\
\Hline
\label{tab2}
\end{tabular}
\end{table*}

\begin{table*}[!t]
\caption{\textcolor{black}{SimGAN architecture details for the refiner/discriminator. Batch normalization is applied both in the refiner/discriminator after each convolutional layer except for the final output layers respectively as in the original paper~\cite{Shrivastava}.}}
\small
  \centering
\begin{tabular}{lcc}
\Hline
\textbf{\textcolor{black}{Refiner}} & \textbf{\textcolor{black}{Activation}} & \textbf{\textcolor{black}{Output Shape}}\\
\hline
\textcolor{black}{Input image} & \textcolor{black}{--} & \textcolor{black}{$\makebox[\widthof{512}][c]{1}\times\makebox[\widthof{1024}][c]{224}\times\makebox[\widthof{1024}][c]{224}$} \\
\textcolor{black}{Conv $9\times9$} & \textcolor{black}{ReLU} & \textcolor{black}{$\makebox[\widthof{512}][c]{64}\times\makebox[\widthof{1024}][c]{224}\times\makebox[\widthof{1024}][c]{224}$} \\
\hline
\textcolor{black}{\blockc{12}} & \textcolor{black}{ReLU} & \textcolor{black}{$\makebox[\widthof{512}][c]{64}\times\makebox[\widthof{1024}][c]{224}\times\makebox[\widthof{1024}][c]{224}$} \\
& \textcolor{black}{--} & \textcolor{black}{$\makebox[\widthof{512}][c]{64}\times\makebox[\widthof{1024}][c]{224}\times\makebox[\widthof{1024}][c]{224}$} \\
\hline
\textcolor{black}{Conv $1\times1$} & \textcolor{black}{Tanh} & \textcolor{black}{$\makebox[\widthof{512}][c]{1}\times\makebox[\widthof{1024}][c]{224}\times\makebox[\widthof{1024}][c]{224}$}  \\
\Hline
\end{tabular}
\hspace{1cm}
\begin{tabular}{lcc}
\Hline
\textcolor{black}{\textbf{Discriminator}} & \textcolor{black}{\textbf{Activation}} & \textcolor{black}{\textbf{Output Shape}}\\
\hline
\textcolor{black}{Input image} & \textcolor{black}{--} & \textcolor{black}{$\makebox[\widthof{512}][c]{1}\times\makebox[\widthof{1024}][c]{224}\times\makebox[\widthof{1024}][c]{224}$} \\
\textcolor{black}{Conv $9\times9$} & \textcolor{black}{ReLU} & \textcolor{black}{$\makebox[\widthof{512}][c]{96}\times\makebox[\widthof{1024}][c]{72}\times\makebox[\widthof{1024}][c]{72}$} \\
\textcolor{black}{Conv $5\times5$} & \textcolor{black}{ReLU} & \textcolor{black}{$\makebox[\widthof{512}][c]{64}\times\makebox[\widthof{1024}][c]{68}\times\makebox[\widthof{1024}][c]{68}$} \\
\textcolor{black}{Maxpool} &  \textcolor{black}{--} & \textcolor{black}{$\makebox[\widthof{512}][c]{64}\times\makebox[\widthof{1024}][c]{34}\times\makebox[\widthof{1024}][c]{34}$} \\
\hline
\textcolor{black}{Conv $5\times5$} & \textcolor{black}{ReLU} & \textcolor{black}{$\makebox[\widthof{512}][c]{64}\times\makebox[\widthof{1024}][c]{15}\times\makebox[\widthof{1024}][c]{15}$} \\
\textcolor{black}{Conv $3\times3$} & \textcolor{black}{ReLU} & \textcolor{black}{$\makebox[\widthof{512}][c]{32}\times\makebox[\widthof{1024}][c]{13}\times\makebox[\widthof{1024}][c]{13}$} \\
\textcolor{black}{Maxpool} & \textcolor{black}{--} & \textcolor{black}{$\makebox[\widthof{512}][c]{32}\times\makebox[\widthof{1024}][c]{7}\times\makebox[\widthof{1024}][c]{7}$} \\
\hline
\textcolor{black}{Conv $1\times1$} & \textcolor{black}{ReLU} & \textcolor{black}{$\makebox[\widthof{512}][c]{32}\times\makebox[\widthof{1024}][c]{7}\times\makebox[\widthof{1024}][c]{7}$} \\
\textcolor{black}{Conv $1\times1$} & \textcolor{black}{ReLU} & \textcolor{black}{$\makebox[\widthof{512}][c]{2}\times\makebox[\widthof{1024}][c]{7}\times\makebox[\widthof{1024}][c]{7}$}  \\
\Hline
\label{tab3}
\end{tabular}
\end{table*}

\vspace{0.1in}

\noindent \textbf{SimGAN}~\cite{Shrivastava} is an image-to-image GAN designed for DA that adopts the self-regularization term/local adversarial loss; it updates a discriminator with a history of refined images.

\vspace{0.1in}

\noindent \textbf{SimGAN Implementation Details}
The SimGAN architecture \textcolor{black}{(i.e., a refiner)} \textcolor{black}{uses} the following loss:

\begin{eqnarray}\label{eq:simgan_loss}
\sum_i  \mathcal{L}_{\text{real}} (\boldsymbol \theta; \mathbf x_i, \mathcal Y ) + \lambda_{\text{reg}} \mathcal{L}_{\text{reg}} (\boldsymbol \theta;  {\mathbf x_i}),
\end{eqnarray}
where \textcolor{black}{$\mathcal{L}(\cdot)$ denotes the loss function,} $\boldsymbol {\theta} $ is the function parameters, $\mathbf x_i$ is the $i^{\text th}$ PGGAN-generated training image, and $\mathcal Y$ is the set of the real images $\mathbf y_j$. The first part $ \mathcal{L}_{\text{real}}$ adds realism to the synthetic images \textcolor{black}{using a discriminator}, while the second part $\mathcal{L}_{\text{reg}}$ preserves the tumor/non-tumor features.


We train \textcolor{black}{the model (Table~\ref{tab3})} for $20,000$ steps with a batch size of 10 and $1.0 \times 10^{-4}$ learning rate for the Stochastic Gradient Descent (SGD) optimizer~\cite{Bottou} \textcolor{black}{without momentum}. The learning rate is reduced by half at 15,000 steps. \textcolor{black}{We train the refiner first with just the self-regularization loss with $\lambda_\text{reg} = 5 \times 10^{-5}$ for $500$ steps; then, for each update of the discriminator, we update the refiner $5$ times. } During training, we apply horizontal flipping as DA.

\subsection{Tumor Detection Using ResNet-50}
\noindent \textbf{Pre-processing.} As ResNet-50's input size is $224 \times 224$ pixels, we resize the whole real images from $240 \times 240$ and whole PGGAN-generated images from $256 \times 256$.

\vspace{0.1in}

\noindent \textbf{ResNet-50}~\cite{He} is a $50$-layer residual learning-based CNN. We adopt it to detect brain tumors in MR images (i.e., the binary classification of \textcolor{black}{tumor/non-tumpor images}) due to its outstanding performance in image classification tasks~\cite{Bianco}\textcolor{black}{, including binary classification~\cite{Yap}}. \textcolor{black}{Chang \textit{et al.}\cite{Chang} also used a similar $34$-layer residual convolutional network for the binary classification of brain tumors (i.e., determining the Isocitrate Dehydrogenase status in low-/high-grade gliomas).}

\textcolor{black}{\noindent \textbf{DA Setups}} To confirm the effect of PGGAN-based DA and its refinement using MUNIT/SimGAN, we compare the following $10$ DA setups under sufficient images both with/without ImageNet~\cite{Russakovsky} pre-training (i.e., 20 DA setups):

\begin{enumerate}
\item $8429$ real images;
\item + $200$k classic DA;
\item + $400$k classic DA;
\item + $200$k PGGAN-based DA;
\item + $200$k PGGAN-based DA w/o clustering/discarding;
\item + $200$k classic DA \& $200$k PGGAN-based DA;
\item + $200$k \textcolor{black}{MUNIT}-refined DA;
\item + $200$k classic DA \& $200$k \textcolor{black}{MUNIT}-refined DA;
\item + $200$k SimGAN-refined DA;
\item + $200$k classic DA \& $200$k SimGAN-refined DA.
\end{enumerate}

\textcolor{black}{Due to the risk of overlooking the tumor diagnosis $via$ medical imaging, higher sensitivity matters much more than higher specificity~\cite{Mazurowski}; thus, we aim to achieve higher sensitivity, using the additional synthetic training images. We perform McNemar’s test on paired tumor detection results~\cite{McNemar} to confirm our two-step GAN-based DA's statistically-significant sensitivity improvement; since this statistical analysis involves multiple comparison tests, we adjust their $p$-values using the Holm–Bonferroni method~\cite{Holm}.}

\begin{figure}[t]
  \centering
  \centerline{\includegraphics[width=\columnwidth]{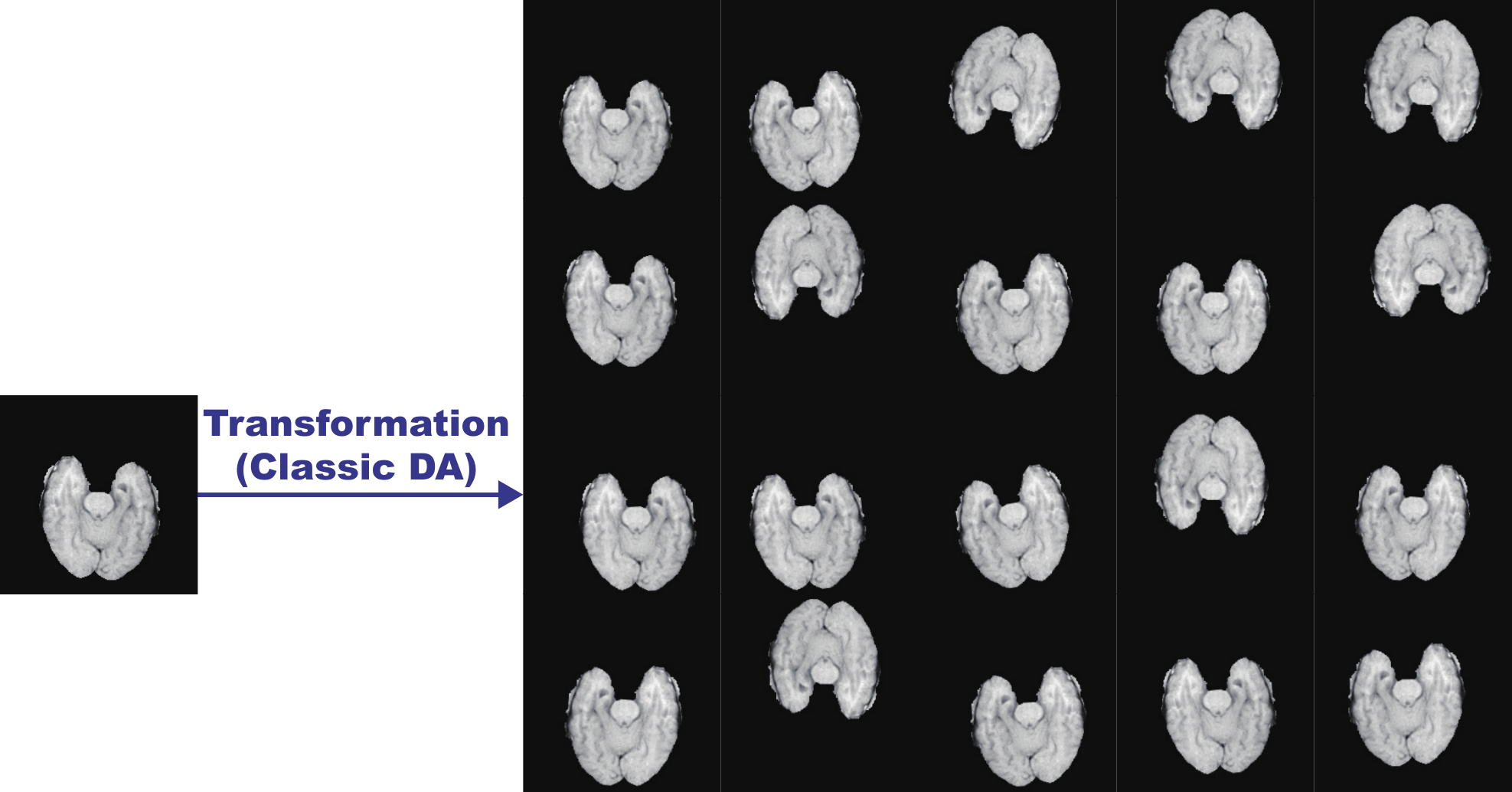}}
\caption{Example real MR image and its geometrically-transformed images.}
\label{fig4}
\end{figure}

\begin{figure}[t]
  \centering
  \centerline{\includegraphics[width=\columnwidth]{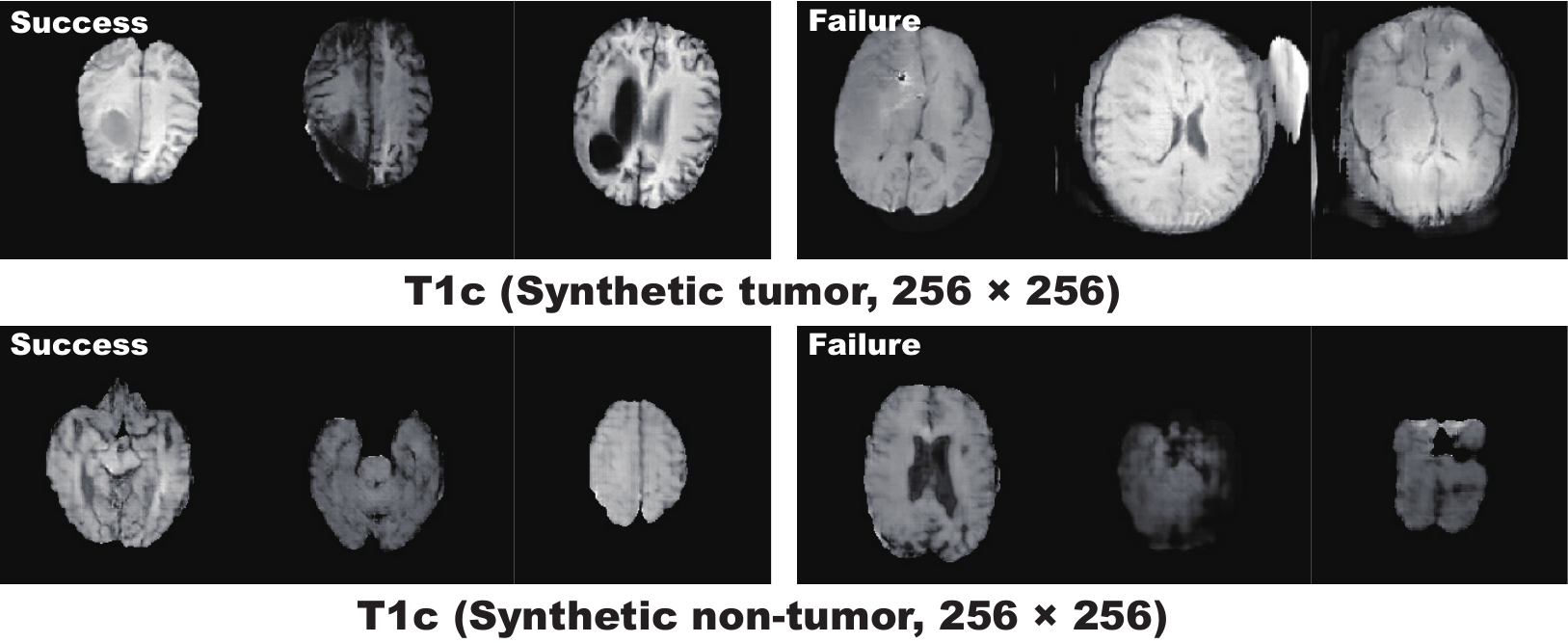}}
\caption{Example PGGAN-generated MR images: (a) Success cases; (b) Failure cases.}
\label{fig5}
\vspace{-0.2in}
\end{figure}

Whereas medical imaging researchers widely use the ImageNet initialization despite different textures of natural/medical images, recent study found that such ImageNet-trained CNNs are biased towards recognizing texture rather than shape~\cite{Geirhos}; thus, we aim to investigate how the medical GAN-based DA affects classification performance with/without the pre-training. As the classic DA, we adopt a random combination of horizontal/vertical flipping, rotation up to $10$ degrees, width/height shift up to $8\%$, shearing up to $8\%$, zooming up to $8\%$, and constant filling of points outside the input boundaries (Fig.~\ref{fig4}). For the PGGAN-based DA and its refinement, we only use success cases after discarding weird-looking synthetic images (Fig.~\ref{fig5}); DenseNet-169~\cite{Iandola} extracts image features and k-means++~\cite{Arthur} clusters the features into $200$ groups, and then we manually discard each cluster containing similar weird-looking images. To verify its effect, we also conduct \textcolor{black}{a} PGGAN-based DA experiment without the discarding step.

\vspace{0.1in}

\noindent \textbf{ResNet-50 Implementation Details}
The ResNet-50 architecture adopts the binary cross-entropy loss for binary classification both with/without ImageNet pre-training. \textcolor{black}{As shown in Table~\ref{tab4},} for robust training, before the final sigmoid layer, we \textcolor{black}{introduce} a $0.5$ dropout~\cite{Srivastava}, linear dense, and batch normalization~\cite{Ioffe} layers---training with GAN-based DA tends to be unstable especially without the batch normalization layer. We use a batch size of $96$, $1.0 \times 10^{-2}$ learning rate for the SGD optimizer~\cite{Bottou} with $0.9$ momentum, and early stopping of $20$ epochs. The learning rate was multiplied by $0.1$ every $20$ epochs for the training from scratch and by $0.5$ every $5$ epochs for the ImageNet pre-training.

\begin{table}[!t]
\caption{\textcolor{black}{ResNet-50 architecture details without/with pre-training. We input grayscale images (i.e., 1 channel) for experiments without pre-training, whereas we input color images (i.e., 3 channels) for experiments with pre-training to use ImageNet initialization. Batch normalization is applied after each convolutional layer as in the original paper~\cite{He}.}}
\small
  \centering
\begin{tabular}{lcc}
\Hline
\textcolor{black}{\textbf{Classifier}} & \textcolor{black}{\textbf{Activation}} & \textcolor{black}{\textbf{Output Shape}}\\
\hline
\textcolor{black}{Input image} & \textcolor{black}{--} & \textcolor{black}{$\makebox[\widthof{512}][c]{1 (3)}\times\makebox[\widthof{1024}][c]{224}\times\makebox[\widthof{1024}][c]{224}$} \\
\textcolor{black}{Conv $7\times7$} & \textcolor{black}{ReLU} & \textcolor{black}{$\makebox[\widthof{512}][c]{64}\times\makebox[\widthof{1024}][c]{112}\times\makebox[\widthof{1024}][c]{112}$} \\
\textcolor{black}{Maxpool} &  \textcolor{black}{--} & \textcolor{black}{$\makebox[\widthof{512}][c]{64}\times\makebox[\widthof{1024}][c]{55}\times\makebox[\widthof{1024}][c]{55}$} \\
\hline
\textcolor{black}{\blockb{3}} & \textcolor{black}{ReLU} & \textcolor{black}{$\makebox[\widthof{512}][c]{64}\times\makebox[\widthof{1024}][c]{55}\times\makebox[\widthof{1024}][c]{55}$} \\
& \textcolor{black}{ReLU} & \textcolor{black}{$\makebox[\widthof{512}][c]{64}\times\makebox[\widthof{1024}][c]{55}\times\makebox[\widthof{1024}][c]{55}$} \\
& \textcolor{black}{ReLU} &\textcolor{black}{$\makebox[\widthof{512}][c]{256}\times\makebox[\widthof{1024}][c]{55}\times\makebox[\widthof{1024}][c]{55}$} \\
\hline
\textcolor{black}{\blockb{4}} & \textcolor{black}{ReLU} & \textcolor{black}{$\makebox[\widthof{512}][c]{128}\times\makebox[\widthof{1024}][c]{28}\times\makebox[\widthof{1024}][c]{28}$} \\
& \textcolor{black}{ReLU} & \textcolor{black}{$\makebox[\widthof{512}][c]{128}\times\makebox[\widthof{1024}][c]{28}\times\makebox[\widthof{1024}][c]{28}$} \\
& \textcolor{black}{ReLU} &\textcolor{black}{$\makebox[\widthof{512}][c]{512}\times\makebox[\widthof{1024}][c]{28}\times\makebox[\widthof{1024}][c]{28}$} \\
\hline
\textcolor{black}{\blockb{6}} & \textcolor{black}{ReLU} & \textcolor{black}{$\makebox[\widthof{512}][c]{256}\times\makebox[\widthof{1024}][c]{14}\times\makebox[\widthof{1024}][c]{14}$} \\
& \textcolor{black}{ReLU} & \textcolor{black}{$\makebox[\widthof{512}][c]{256}\times\makebox[\widthof{1024}][c]{14}\times\makebox[\widthof{1024}][c]{14}$} \\
& \textcolor{black}{ReLU} &\textcolor{black}{$\makebox[\widthof{512}][c]{1024}\times\makebox[\widthof{1024}][c]{14}\times\makebox[\widthof{1024}][c]{14}$} \\
\hline
\textcolor{black}{\blockb{3}} & \textcolor{black}{ReLU} & \textcolor{black}{$\makebox[\widthof{512}][c]{512}\times\makebox[\widthof{1024}][c]{7}\times\makebox[\widthof{1024}][c]{7}$} \\
& \textcolor{black}{ReLU} & \textcolor{black}{$\makebox[\widthof{512}][c]{512}\times\makebox[\widthof{1024}][c]{7}\times\makebox[\widthof{1024}][c]{7}$} \\
& \textcolor{black}{ReLU} &\textcolor{black}{$\makebox[\widthof{512}][c]{2048}\times\makebox[\widthof{1024}][c]{7}\times\makebox[\widthof{1024}][c]{7}$} \\
\hline
\textcolor{black}{AveragePool} & \textcolor{black}{--} & \textcolor{black}{$\makebox[\widthof{512}][c]{2048}\times\makebox[\widthof{1024}][c]{1}\times\makebox[\widthof{1024}][c]{1}$}  \\
\textcolor{black}{Flatten} & \textcolor{black}{--} & \textcolor{black}{$\makebox[\widthof{1024}][c]{2048}$}  \\
\textcolor{black}{$0.5$ Dropout} & \textcolor{black}{--} & \textcolor{black}{$\makebox[\widthof{1024}][c]{2048}$}  \\
\textcolor{black}{Dense} & \textcolor{black}{--} & \textcolor{black}{$\makebox[\widthof{1024}][c]{2}$}  \\
\textcolor{black}{BatchNorm} & \textcolor{black}{Sigmoid} & \textcolor{black}{$\makebox[\widthof{1024}][c]{2}$}  \\
\Hline
\label{tab4}
\end{tabular}
\end{table}

\begin{table*}[!t]
\caption{ResNet-50 tumor detection (i.e., binary classification) results with various DA, with (without) ImageNet pre-training. \textcolor{black}{Sensitivity and specificity consider the slight tumor/non-tumor class imbalance (about 6:5) in the test set. Boldface indicates the best performance.}}
\centering
\begin{small}
\begin{tabular}{lrrr}
\Hline\noalign{\smallskip}
\multicolumn{1}{l}{\bfseries DA Setups}  & \multicolumn{1}{c}{\bfseries Accuracy} \hspace{-0.1in} (\%) & \multicolumn{1}{c}{\bfseries Sensitivity} \hspace{-0.1in} (\%) & \multicolumn{1}{c}{\bfseries Specificity} \hspace{-0.1in} (\%)  \\\noalign{\smallskip}\hline\noalign{\smallskip}
\textcolor{black}{(1)} 8,429 real images & \textcolor{black}{93.14} (\textcolor{black}{86.33}) \hspace{-0.1in} & \textcolor{black}{90.91} (\textcolor{black}{88.88}) \hspace{-0.1in} & \textcolor{black}{95.85} (\textcolor{black}{83.22}) \hspace{-0.1in} \\
\textcolor{black}{(2)} + 200k classic DA & \textcolor{black}{95.01} (\textcolor{black}{92.20}) \hspace{-0.1in} & \textcolor{black}{93.67} (\textcolor{black}{89.94}) \hspace{-0.1in} & \textcolor{black}{96.64} (\textcolor{black}{94.97}) \hspace{-0.1in} \\
\textcolor{black}{(3)} + 400k classic DA & \textcolor{black}{94.83} (\textcolor{black}{93.23}) \hspace{-0.1in} & \textcolor{black}{91.88} (\textcolor{black}{90.91}) \hspace{-0.1in} & \textcolor{black}{98.42} (\textcolor{black}{96.05}) \hspace{-0.1in} \\\noalign{\smallskip}\hline\noalign{\smallskip}
\textcolor{black}{(4)} + 200k PGGAN-based DA & \textcolor{black}{93.94} (\textcolor{black}{86.19}) \hspace{-0.1in} & \textcolor{black}{92.61} (\textcolor{black}{87.26}) \hspace{-0.1in} & \textcolor{black}{95.56} (\textcolor{black}{84.90}) \hspace{-0.1in} \\
\textcolor{black}{(5)} + 200k PGGAN-based DA w/o clustering/discarding & \textcolor{black}{94.83} (\textcolor{black}{80.67}) \hspace{-0.1in} & \textcolor{black}{91.88} (\textcolor{black}{80.19}) \hspace{-0.1in} & \textcolor{black}{98.42} (\textcolor{black}{81.24}) \hspace{-0.1in} \\
\textcolor{black}{(6)} + 200k classic DA \& 200k PGGAN-based DA & \textcolor{black}{96.17} (\textcolor{black}{95.59}) \hspace{-0.1in} & \textcolor{black}{93.99} (\textcolor{black}{94.16}) \hspace{-0.1in} & \textbf{\textcolor{black}{98.82}} (\textcolor{black}{97.33}) \hspace{-0.1in} \\\noalign{\smallskip}\hline\noalign{\smallskip}
\textcolor{black}{(7)} + 200k \textcolor{black}{MUNIT}-refined DA & \textcolor{black}{94.30} (\textcolor{black}{83.65}) \hspace{-0.1in} & \textcolor{black}{93.02} (\textcolor{black}{87.82}) \hspace{-0.1in} & \textcolor{black}{95.85} (\textcolor{black}{78.58}) \hspace{-0.1in} \\
\textcolor{black}{(8)} + 200k classic DA \& 200k \textcolor{black}{MUNIT}-refined DA & \textbf{\textcolor{black}{96.70}} (\textcolor{black}{96.35}) \hspace{-0.1in} & \textcolor{black}{95.45} (\textbf{\textcolor{black}{97.48}}) \hspace{-0.1in} & \textcolor{black}{98.22} (\textcolor{black}{94.97}) \hspace{-0.1in} \\\noalign{\smallskip}\hline\noalign{\smallskip}
\textcolor{black}{(9)} + 200k SimGAN-refined DA & \textcolor{black}{94.48} (\textcolor{black}{77.64}) \hspace{-0.1in} & \textcolor{black}{92.29} (\textcolor{black}{82.31}) \hspace{-0.1in} & \textcolor{black}{97.14} (\textcolor{black}{71.96}) \hspace{-0.1in} \\
\textcolor{black}{(10)} + 200k classic DA \& 200k SimGAN-refined DA & \textcolor{black}{96.39} (\textcolor{black}{95.01}) \hspace{-0.1in} & \textcolor{black}{95.13} (\textcolor{black}{95.05}) \hspace{-0.1in} & \textcolor{black}{97.93} (\textcolor{black}{94.97}) \hspace{-0.1in} \\

\noalign{\smallskip}\Hline\noalign{\smallskip}
\label{tab5}
\end{tabular}
\end{small}
\vspace{-0.1in}
\end{table*}

\begin{table*}[!t]
\caption{\textcolor{black}{McNemar's test $p$-values for the pairwise comparison of the ResNet-50 tumor detection results in terms of accuracy, sensitivity, specificity, respectively. We compare our two-step GAN-based DA setups and all the other DA setups. All numbers within parentheses refer to DA setups on Table~\ref{tab5} and PT denotes pre-training. Boldface indicates statistical significance (threshold $p$-value $< 0.05$).}}
\centering
\begin{small}
\scalebox{0.755}{
\begin{tabular}{lrrrlrrrlrrr}
\Hline\noalign{\smallskip}
\multicolumn{1}{l}{\bfseries DA Setup Comparison} & \multicolumn{1}{c}{\bfseries Accu} \hspace{-0.1in} & \multicolumn{1}{c}{\bfseries Sens} \hspace{-0.1in}  & \multicolumn{1}{c}{\bfseries Spec} \hspace{0.0in} & \multicolumn{1}{l}{\bfseries DA Setup Comparison} & \multicolumn{1}{c}{\bfseries Accu} \hspace{-0.1in} & \multicolumn{1}{c}{\bfseries Sens} \hspace{-0.1in}  & \multicolumn{1}{c}{\bfseries Spec} \hspace{0.0in} & \multicolumn{1}{l}{\bfseries DA Setup Comparison} & \multicolumn{1}{c}{\bfseries Accu} \hspace{-0.1in} & \multicolumn{1}{c}{\bfseries Sens} \hspace{-0.1in}  & \multicolumn{1}{c}{\bfseries Spec} \hspace{-0.1in} \\\noalign{\smallskip}\hline\noalign{\smallskip}
\textcolor{black}{(7) w/ PT \textit{vs} (1) w/ PT} & \textcolor{black}{0.693} \hspace{-0.1in} & \textcolor{black}{0.206} \hspace{-0.1in} & \textcolor{black}{1} \hspace{0.0in} &

\textcolor{black}{(7) w/ PT \textit{vs} (1) w/o PT} & \textcolor{black}{\textbf{$<$ 0.001}} \hspace{-0.1in} & \textcolor{black}{\textbf{0.002}} \hspace{-0.1in} & \textcolor{black}{\textbf{$<$ 0.001}} \hspace{0.0in} &

\textcolor{black}{(7) w/ PT \textit{vs} (2) w/ PT} & \textcolor{black}{1} \hspace{-0.1in} & \textcolor{black}{1} \hspace{-0.1in} & \textcolor{black}{1} \hspace{-0.1in}
\\\noalign{\smallskip}\hline\noalign{\smallskip}

\textcolor{black}{(7) w/ PT \textit{vs} (2) w/o PT} & \textcolor{black}{\textbf{0.034}} \hspace{-0.1in} & \textcolor{black}{\textbf{0.024}} \hspace{-0.1in} & \textcolor{black}{1} \hspace{0.0in} &

\textcolor{black}{(7) w/ PT \textit{vs} (3) w/ PT} & \textcolor{black}{1} \hspace{-0.1in} & \textcolor{black}{1} \hspace{-0.1in} & \textcolor{black}{\textbf{0.035}} \hspace{0.0in} &

\textcolor{black}{(7) w/ PT \textit{vs} (3) w/o PT} & \textcolor{black}{1} \hspace{-0.1in} & \textcolor{black}{0.468} \hspace{-0.1in} & \textcolor{black}{1} \hspace{-0.1in}
\\\noalign{\smallskip}\hline\noalign{\smallskip}

\textcolor{black}{(7) w/ PT \textit{vs} (4) w/ PT} & \textcolor{black}{1} \hspace{-0.1in} & \textcolor{black}{1} \hspace{-0.1in} & \textcolor{black}{1} \hspace{0.0in} &

\textcolor{black}{(7) w/ PT \textit{vs} (4) w/o PT} & \textcolor{black}{\textbf{$<$ 0.001}} \hspace{-0.1in} & \textcolor{black}{\textbf{$<$ 0.001}} \hspace{-0.1in} & \textcolor{black}{\textbf{$<$ 0.001}} \hspace{0.0in} &

\textcolor{black}{(7) w/ PT \textit{vs} (5) w/ PT} & \textcolor{black}{1} \hspace{-0.1in} & \textcolor{black}{1} \hspace{-0.1in} & \textcolor{black}{\textbf{0.003}} \hspace{-0.1in}
\\\noalign{\smallskip}\hline\noalign{\smallskip}

\textcolor{black}{(7) w/ PT \textit{vs} (5) w/o PT} & \textcolor{black}{\textbf{$<$ 0.001}} \hspace{-0.1in} & \textcolor{black}{\textbf{$<$ 0.001}} \hspace{-0.1in} & \textcolor{black}{\textbf{$<$ 0.001}} \hspace{0.0in} &

\textcolor{black}{(7) w/ PT \textit{vs} (6) w/ PT} & \textcolor{black}{\textbf{0.009}} \hspace{-0.1in} & \textcolor{black}{1} \hspace{-0.1in} & \textcolor{black}{\textbf{$<$ 0.001}} \hspace{0.0in} &

\textcolor{black}{(7) w/ PT \textit{vs} (6) w/o PT} & \textcolor{black}{0.397} \hspace{-0.1in} & \textcolor{black}{1} \hspace{-0.1in} & \textcolor{black}{1} \hspace{-0.1in}
\\\noalign{\smallskip}\hline\noalign{\smallskip}

\textcolor{black}{(7) w/ PT \textit{vs} (7) w/o PT} & \textcolor{black}{\textbf{$<$ 0.001}} \hspace{-0.1in} & \textcolor{black}{\textbf{$<$ 0.001}} \hspace{-0.1in} & \textcolor{black}{\textbf{$<$ 0.001}} \hspace{0.0in} &

\textcolor{black}{(7) w/ PT \textit{vs} (8) w/ PT} & \textcolor{black}{\textbf{$<$ 0.001}} \hspace{-0.1in} & \textcolor{black}{\textbf{0.025}} \hspace{-0.1in} & \textcolor{black}{\textbf{0.045}} \hspace{0.0in} &

\textcolor{black}{(7) w/ PT \textit{vs} (8) w/o PT} & \textcolor{black}{\textbf{0.008}} \hspace{-0.1in} & \textcolor{black}{\textbf{$<$ 0.001}} \hspace{-0.1in} & \textcolor{black}{1} \hspace{-0.1in}
\\\noalign{\smallskip}\hline\noalign{\smallskip}

\textcolor{black}{(7) w/ PT \textit{vs} (9) w/ PT} & \textcolor{black}{1} \hspace{-0.1in} & \textcolor{black}{1} \hspace{-0.1in} & \textcolor{black}{1} \hspace{0.0in} &

\textcolor{black}{(7) w/ PT \textit{vs} (9) w/o PT} & \textcolor{black}{\textbf{$<$ 0.001}} \hspace{-0.1in} & \textcolor{black}{\textbf{$<$ 0.001}} \hspace{-0.1in} & \textcolor{black}{\textbf{$<$ 0.001}} \hspace{0.0in} &

\textcolor{black}{(7) w/ PT \textit{vs} (10) w/ PT} & \textcolor{black}{\textbf{$<$ 0.001}} \hspace{-0.1in} & \textcolor{black}{0.077} \hspace{-0.1in} & \textcolor{black}{0.108} \hspace{-0.1in}
\\\noalign{\smallskip}\hline\noalign{\smallskip}

\textcolor{black}{(7) w/ PT \textit{vs} (10) w/o PT} & \textcolor{black}{1} \hspace{-0.1in} & \textcolor{black}{0.206} \hspace{-0.1in} & \textcolor{black}{1} \hspace{0.0in} &

\textcolor{black}{(7) w/o PT \textit{vs} (1) w/ PT} & \textcolor{black}{\textbf{$<$ 0.001}} \hspace{-0.1in} & \textcolor{black}{0.135} \hspace{-0.1in} & \textcolor{black}{\textbf{$<$ 0.001}} \hspace{0.0in} &

\textcolor{black}{(7) w/o PT \textit{vs} (1) w/o PT} & \textcolor{black}{\textbf{0.026}} \hspace{-0.1in} & \textcolor{black}{1} \hspace{-0.1in} & \textcolor{black}{\textbf{0.014}} \hspace{-0.1in}
\\\noalign{\smallskip}\hline\noalign{\smallskip}

\textcolor{black}{(7) w/o PT \textit{vs} (2) w/ PT} & \textcolor{black}{\textbf{$<$ 0.001}} \hspace{-0.1in} & \textcolor{black}{\textbf{$<$ 0.001}} \hspace{-0.1in} & \textcolor{black}{\textbf{$<$ 0.001}} \hspace{0.0in} &

\textcolor{black}{(7) w/o PT \textit{vs} (2) w/o PT} & \textcolor{black}{\textbf{$<$ 0.001}} \hspace{-0.1in} & \textcolor{black}{1} \hspace{-0.1in} & \textcolor{black}{\textbf{$<$ 0.001}} \hspace{0.0in} &

\textcolor{black}{(7) w/o PT \textit{vs} (3) w/ PT} & \textcolor{black}{\textbf{$<$ 0.001}} \hspace{-0.1in} & \textcolor{black}{\textbf{0.020}} \hspace{-0.1in} & \textcolor{black}{\textbf{$<$ 0.001}} \hspace{-0.1in}
\\\noalign{\smallskip}\hline\noalign{\smallskip}

\textcolor{black}{(7) w/o PT \textit{vs} (3) w/o PT} & \textcolor{black}{\textbf{$<$ 0.001}} \hspace{-0.1in} & \textcolor{black}{0.147} \hspace{-0.1in} & \textcolor{black}{\textbf{$<$ 0.001}} \hspace{0.0in} &

\textcolor{black}{(7) w/o PT \textit{vs} (4) w/ PT} & \textcolor{black}{\textbf{$<$ 0.001}} \hspace{-0.1in} & \textcolor{black}{\textbf{0.002}} \hspace{-0.1in} & \textcolor{black}{\textbf{$<$ 0.001}} \hspace{0.0in} &

\textcolor{black}{(7) w/o PT \textit{vs} (4) w/o PT} & \textcolor{black}{\textbf{0.044}} \hspace{-0.1in} & \textcolor{black}{1} \hspace{-0.1in} & \textcolor{black}{\textbf{$<$ 0.001}} \hspace{-0.1in}
\\\noalign{\smallskip}\hline\noalign{\smallskip}

\textcolor{black}{(7) w/o PT \textit{vs} (5) w/ PT} & \textcolor{black}{\textbf{$<$ 0.001}} \hspace{-0.1in} & \textcolor{black}{\textbf{0.015}} \hspace{-0.1in} & \textcolor{black}{\textbf{$<$ 0.001}} \hspace{0.0in} &

\textcolor{black}{(7) w/o PT \textit{vs} (5) w/o PT} & \textcolor{black}{\textbf{0.011}} \hspace{-0.1in} & \textcolor{black}{\textbf{$<$ 0.001}} \hspace{-0.1in} & \textcolor{black}{1} \hspace{0.0in} &

\textcolor{black}{(7) w/o PT \textit{vs} (6) w/ PT} & \textcolor{black}{\textbf{$<$ 0.001}} \hspace{-0.1in} & \textcolor{black}{\textbf{$<$ 0.001}} \hspace{-0.1in} & \textcolor{black}{\textbf{$<$ 0.001}} \hspace{-0.1in}
\\\noalign{\smallskip}\hline\noalign{\smallskip}

\textcolor{black}{(7) w/o PT \textit{vs} (6) w/o PT} & \textcolor{black}{\textbf{$<$ 0.001}} \hspace{-0.1in} & \textcolor{black}{\textbf{$<$ 0.001}} \hspace{-0.1in} & \textcolor{black}{\textbf{$<$ 0.001}} \hspace{0.0in} &

\textcolor{black}{(7) w/o PT \textit{vs} (8) w/ PT} & \textcolor{black}{\textbf{$<$ 0.001}} \hspace{-0.1in} & \textcolor{black}{\textbf{$<$ 0.001}} \hspace{-0.1in} & \textcolor{black}{\textbf{$<$ 0.001}} \hspace{0.0in} &

\textcolor{black}{(7) w/o PT \textit{vs} (8) w/o PT} & \textcolor{black}{\textbf{$<$ 0.001}} \hspace{-0.1in} & \textcolor{black}{\textbf{$<$ 0.001}} \hspace{-0.1in} & \textcolor{black}{\textbf{$<$ 0.001}} \hspace{-0.1in}
\\\noalign{\smallskip}\hline\noalign{\smallskip}

\textcolor{black}{(7) w/o PT \textit{vs} (9) w/ PT} & \textcolor{black}{\textbf{$<$ 0.001}} \hspace{-0.1in} & \textcolor{black}{\textbf{0.004}} \hspace{-0.1in} & \textcolor{black}{\textbf{$<$ 0.001}} \hspace{0.0in} &

\textcolor{black}{(7) w/o PT \textit{vs} (9) w/o PT} & \textcolor{black}{\textbf{$<$ 0.001}} \hspace{-0.1in} & \textcolor{black}{\textbf{$<$ 0.001}} \hspace{-0.1in} & \textcolor{black}{\textbf{$<$ 0.001}} \hspace{0.0in} &

\textcolor{black}{(7) w/o PT \textit{vs} (10) w/ PT} & \textcolor{black}{\textbf{$<$ 0.001}} \hspace{-0.1in} & \textcolor{black}{\textbf{$<$ 0.001}} \hspace{-0.1in} & \textcolor{black}{\textbf{$<$ 0.001}} \hspace{-0.1in}
\\\noalign{\smallskip}\hline\noalign{\smallskip}

\textcolor{black}{(7) w/o PT \textit{vs} (10) w/o PT} & \textcolor{black}{\textbf{$<$ 0.001}} \hspace{-0.1in} & \textcolor{black}{\textbf{$<$ 0.001}} \hspace{-0.1in} & \textcolor{black}{\textbf{$<$ 0.001}} \hspace{0.0in} &

\textcolor{black}{(8) w/ PT \textit{vs} (1) w PT} & \textcolor{black}{\textbf{$<$ 0.001}} \hspace{-0.1in} & \textcolor{black}{\textbf{$<$ 0.001}} \hspace{-0.1in} & \textcolor{black}{\textbf{0.010}} \hspace{0.0in} &

\textcolor{black}{(8) w/ PT \textit{vs} (1) w/o PT} & \textcolor{black}{\textbf{$<$ 0.001}} \hspace{-0.1in} & \textcolor{black}{\textbf{$<$ 0.001}} \hspace{-0.1in} & \textcolor{black}{\textbf{$<$ 0.001}} \hspace{-0.1in}
\\\noalign{\smallskip}\hline\noalign{\smallskip}

\textcolor{black}{(8) w/ PT \textit{vs} (2) w/ PT} & \textcolor{black}{\textbf{$<$ 0.001}} \hspace{-0.1in} & \textcolor{black}{0.074} \hspace{-0.1in} & \textcolor{black}{0.206} \hspace{0.0in} &

\textcolor{black}{(8) w/ PT \textit{vs} (2) w/o PT} & \textcolor{black}{\textbf{$<$ 0.001}} \hspace{-0.1in} & \textcolor{black}{\textbf{$<$ 0.001}} \hspace{-0.1in} & \textcolor{black}{\textbf{$<$ 0.001}} \hspace{0.0in} &

\textcolor{black}{(8) w/ PT \textit{vs} (3) w/ PT} & \textcolor{black}{\textbf{0.002}} \hspace{-0.1in} & \textcolor{black}{\textbf{$<$ 0.001}} \hspace{-0.1in} & \textcolor{black}{1} \hspace{-0.1in}
\\\noalign{\smallskip}\hline\noalign{\smallskip}

\textcolor{black}{(8) w/ PT \textit{vs} (3) w/o PT} & \textcolor{black}{\textbf{$<$ 0.001}} \hspace{-0.1in} & \textcolor{black}{\textbf{$<$ 0.001}} \hspace{-0.1in} & \textcolor{black}{0.112} \hspace{0.0in} &

\textcolor{black}{(8) w/ PT \textit{vs} (4) w/ PT} & \textcolor{black}{\textbf{$<$ 0.001}} \hspace{-0.1in} & \textcolor{black}{\textbf{$<$ 0.001}} \hspace{-0.1in} & \textcolor{black}{\textbf{0.006}} \hspace{0.0in} &

\textcolor{black}{(8) w/ PT \textit{vs} (4) w/o PT} & \textcolor{black}{\textbf{$<$ 0.001}} \hspace{-0.1in} & \textcolor{black}{\textbf{$<$ 0.001}} \hspace{-0.1in} & \textcolor{black}{\textbf{$<$ 0.001}} \hspace{-0.1in}
\\\noalign{\smallskip}\hline\noalign{\smallskip}

\textcolor{black}{(8) w/ PT \textit{vs} (5) w/ PT} & \textcolor{black}{\textbf{0.002}} \hspace{-0.1in} & \textcolor{black}{\textbf{$<$ 0.001}} \hspace{-0.1in} & \textcolor{black}{1} \hspace{0.0in} &

\textcolor{black}{(8) w/ PT \textit{vs} (5) w/o PT} & \textcolor{black}{\textbf{$<$ 0.001}} \hspace{-0.1in} & \textcolor{black}{\textbf{$<$ 0.001}} \hspace{-0.1in} & \textcolor{black}{\textbf{$<$ 0.001}} \hspace{0.0in} &

\textcolor{black}{(8) w/ PT \textit{vs} (6) w/ PT} & \textcolor{black}{1} \hspace{-0.1in} & \textcolor{black}{0.128} \hspace{-0.1in} & \textcolor{black}{1} \hspace{-0.1in}
\\\noalign{\smallskip}\hline\noalign{\smallskip}

\textcolor{black}{(8) w/ PT \textit{vs} (6) w/o PT} & \textcolor{black}{0.222} \hspace{-0.1in} & \textcolor{black}{0.760} \hspace{-0.1in} & \textcolor{black}{1} \hspace{0.0in} &

\textcolor{black}{(8) w/ PT \textit{vs} (8) w/o PT} & \textcolor{black}{1} \hspace{-0.1in} & \textcolor{black}{\textbf{0.008}} \hspace{-0.1in} & \textcolor{black}{\textbf{$<$ 0.001}} \hspace{0.0in} &

\textcolor{black}{(8) w/ PT \textit{vs} (9) w/ PT} & \textcolor{black}{\textbf{$<$ 0.001}} \hspace{-0.1in} & \textcolor{black}{\textbf{$<$ 0.001}} \hspace{-0.1in} & \textcolor{black}{1} \hspace{-0.1in}
\\\noalign{\smallskip}\hline\noalign{\smallskip}

\textcolor{black}{(8) w/ PT \textit{vs} (9) w/o PT} & \textcolor{black}{\textbf{$<$ 0.001}} \hspace{-0.1in} & \textcolor{black}{\textbf{$<$ 0.001}} \hspace{-0.1in} & \textcolor{black}{\textbf{$<$ 0.001}} \hspace{0.0in} &

\textcolor{black}{(8) w/ PT \textit{vs} (10) w/ PT} & \textcolor{black}{1} \hspace{-0.1in} & \textcolor{black}{1} \hspace{-0.1in} & \textcolor{black}{1} \hspace{0.0in} &

\textcolor{black}{(8) w/ PT \textit{vs} (10) w/o PT} & \textcolor{black}{\textbf{0.007}} \hspace{-0.1in} & \textcolor{black}{1} \hspace{-0.1in} & \textcolor{black}{0} \hspace{-0.1in}
\\\noalign{\smallskip}\hline\noalign{\smallskip}

\textcolor{black}{(8) w/o PT \textit{vs} (1) w/ PT} & \textcolor{black}{\textbf{$<$ 0.001}} \hspace{-0.1in} & \textcolor{black}{\textbf{$<$ 0.001}} \hspace{-0.1in} & \textcolor{black}{1} \hspace{0.0in} &

\textcolor{black}{(8) w/o PT \textit{vs} (1) w/o PT} & \textcolor{black}{\textbf{$<$ 0.001}} \hspace{-0.1in} & \textcolor{black}{\textbf{$<$ 0.001}} \hspace{-0.1in} & \textcolor{black}{\textbf{$<$ 0.001}} \hspace{0.0in} &

\textcolor{black}{(8) w/o PT \textit{vs} (2) w/ PT} & \textcolor{black}{0.179} \hspace{-0.1in} & \textcolor{black}{\textbf{$<$ 0.001}} \hspace{-0.1in} & \textcolor{black}{0.588} \hspace{-0.1in}
\\\noalign{\smallskip}\hline\noalign{\smallskip}

\textcolor{black}{(8) w/o PT \textit{vs} (2) w/o PT} & \textcolor{black}{\textbf{$<$ 0.001}} \hspace{-0.1in} & \textcolor{black}{\textbf{$<$ 0.001}} \hspace{-0.1in} & \textcolor{black}{1} \hspace{0.0in} &

\textcolor{black}{(8) w/o PT \textit{vs} (3) w/ PT} & \textcolor{black}{0.101} \hspace{-0.1in} & \textcolor{black}{\textbf{$<$ 0.001}} \hspace{-0.1in} & \textcolor{black}{\textbf{$<$ 0.001}} \hspace{0.0in} &

\textcolor{black}{(8) w/o PT \textit{vs} (3) w/o PT} & \textcolor{black}{\textbf{$<$ 0.001}} \hspace{-0.1in} & \textcolor{black}{\textbf{$<$ 0.001}} \hspace{-0.1in} & \textcolor{black}{1} \hspace{-0.1in}
\\\noalign{\smallskip}\hline\noalign{\smallskip}

\textcolor{black}{(8) w/o PT \textit{vs} (4) w/ PT} & \textcolor{black}{\textbf{$<$ 0.001}} \hspace{-0.1in} & \textcolor{black}{\textbf{$<$ 0.001}} \hspace{-0.1in} & \textcolor{black}{1} \hspace{0.0in} &

\textcolor{black}{(8) w/o PT \textit{vs} (4) w/o PT} & \textcolor{black}{\textbf{$<$ 0.001}} \hspace{-0.1in} & \textcolor{black}{\textbf{$<$ 0.001}} \hspace{-0.1in} & \textcolor{black}{\textbf{$<$ 0.001}} \hspace{0.0in} &

\textcolor{black}{(8) w/o PT \textit{vs} (5) w/ PT} & \textcolor{black}{0.197} \hspace{-0.1in} & \textcolor{black}{\textbf{$<$ 0.001}} \hspace{-0.1in} & \textcolor{black}{\textbf{$<$ 0.001}} \hspace{-0.1in}
\\\noalign{\smallskip}\hline\noalign{\smallskip}

\textcolor{black}{(8) w/o PT \textit{vs} (5) w/o PT} & \textcolor{black}{\textbf{$<$ 0.001}} \hspace{-0.1in} & \textcolor{black}{\textbf{$<$ 0.001}} \hspace{-0.1in} & \textcolor{black}{\textbf{$<$ 0.001}} \hspace{0.0in} &

\textcolor{black}{(8) w/o PT \textit{vs} (6) w/ PT} & \textcolor{black}{1} \hspace{-0.1in} & \textcolor{black}{\textbf{$<$ 0.001}} \hspace{-0.1in} & \textcolor{black}{\textbf{$<$ 0.001}} \hspace{0.0in} &

\textcolor{black}{(8) w/o PT \textit{vs} (6) w/o PT} & \textcolor{black}{1} \hspace{-0.1in} & \textcolor{black}{\textbf{$<$ 0.001}} \hspace{-0.1in} & \textcolor{black}{\textbf{0.007}} \hspace{-0.1in}
\\\noalign{\smallskip}\hline\noalign{\smallskip}

\textcolor{black}{(8) w/o PT \textit{vs} (9) w/ PT} & \textcolor{black}{\textbf{0.023}} \hspace{-0.1in} & \textcolor{black}{\textbf{$<$ 0.001}} \hspace{-0.1in} & \textcolor{black}{0.256} \hspace{0.0in} &

\textcolor{black}{(8) w/o PT \textit{vs} (9) w/o PT} & \textcolor{black}{\textbf{$<$ 0.001}} \hspace{-0.1in} & \textcolor{black}{\textbf{$<$ 0.001}} \hspace{-0.1in} & \textcolor{black}{\textbf{$<$ 0.001}} \hspace{0.0in} &

\textcolor{black}{(8) w/o PT \textit{vs} (10) w/ PT} & \textcolor{black}{1} \hspace{-0.1in} & \textcolor{black}{\textbf{0.002}} \hspace{-0.1in} & \textcolor{black}{\textbf{$<$ 0.001}} \hspace{-0.1in}
\\\noalign{\smallskip}\hline\noalign{\smallskip}

\textcolor{black}{(8) w/o PT \textit{vs} (10) w/o PT} & \textcolor{black}{0.143} \hspace{-0.1in} & \textcolor{black}{\textbf{0.005}} \hspace{-0.1in} & \textcolor{black}{1} \hspace{0.0in} &

\textcolor{black}{(9) w/ PT \textit{vs} (1) w/ PT} & \textcolor{black}{0.387} \hspace{-0.1in} & \textcolor{black}{1} \hspace{-0.1in} & \textcolor{black}{1} \hspace{0.0in} &

\textcolor{black}{(9) w/ PT \textit{vs} (1) w/o PT} & \textcolor{black}{\textbf{$<$ 0.001}} \hspace{-0.1in} & \textcolor{black}{\textbf{0.046}} \hspace{-0.1in} & \textcolor{black}{\textbf{$<$ 0.001}} \hspace{-0.1in}
\\\noalign{\smallskip}\hline\noalign{\smallskip}

\textcolor{black}{(9) w/ PT \textit{vs} (2) w/ PT} & \textcolor{black}{1} \hspace{-0.1in} & \textcolor{black}{1} \hspace{-0.1in} & \textcolor{black}{1} \hspace{0.0in} &

\textcolor{black}{(9) w/ PT \textit{vs} (2) w/o PT} & \textcolor{black}{\textbf{0.008}} \hspace{-0.1in} & \textcolor{black}{0.262} \hspace{-0.1in} & \textcolor{black}{0.321} \hspace{0.0in} &

\textcolor{black}{(9) w/ PT \textit{vs} (3) w/ PT} & \textcolor{black}{1} \hspace{-0.1in} & \textcolor{black}{1} \hspace{-0.1in} & \textcolor{black}{0.931} \hspace{-0.1in}
\\\noalign{\smallskip}\hline\noalign{\smallskip}

\textcolor{black}{(9) w/ PT \textit{vs} (3) w/o PT} & \textcolor{black}{0.910} \hspace{-0.1in} & \textcolor{black}{1} \hspace{-0.1in} & \textcolor{black}{1} \hspace{0.0in} &

\textcolor{black}{(9) w/ PT \textit{vs} (4) w/ PT} & \textcolor{black}{1} \hspace{-0.1in} & \textcolor{black}{1} \hspace{-0.1in} & \textcolor{black}{0.764} \hspace{0.0in} &

\textcolor{black}{(9) w/ PT \textit{vs} (4) w/o PT} & \textcolor{black}{\textbf{$<$ 0.001}} \hspace{-0.1in} & \textcolor{black}{\textbf{$<$ 0.001}} \hspace{-0.1in} & \textcolor{black}{\textbf{$<$ 0.001}} \hspace{-0.1in}
\\\noalign{\smallskip}\hline\noalign{\smallskip}

\textcolor{black}{(9) w/ PT \textit{vs} (5) w/ PT} & \textcolor{black}{1} \hspace{-0.1in} & \textcolor{black}{1} \hspace{-0.1in} & \textcolor{black}{0.639} \hspace{0.0in} &

\textcolor{black}{(9) w/ PT \textit{vs} (5) w/o PT} & \textcolor{black}{\textbf{$<$ 0.001}} \hspace{-0.1in} & \textcolor{black}{\textbf{$<$ 0.001}} \hspace{-0.1in} & \textcolor{black}{\textbf{$<$ 0.001}} \hspace{0.0in} &

\textcolor{black}{(9) w/ PT \textit{vs} (6) w/ PT} & \textcolor{black}{\textbf{0.014}} \hspace{-0.1in} & \textcolor{black}{0.660} \hspace{-0.1in} & \textcolor{black}{0.066} \hspace{-0.1in}
\\\noalign{\smallskip}\hline\noalign{\smallskip}

\textcolor{black}{(9) w/ PT \textit{vs} (6) w/o PT} & \textcolor{black}{0.716} \hspace{-0.1in} & \textcolor{black}{0.365} \hspace{-0.1in} & \textcolor{black}{1} \hspace{0.0in} &

\textcolor{black}{(9) w/ PT \textit{vs} (9) w/o PT} & \textcolor{black}{\textbf{$<$ 0.001}} \hspace{-0.1in} & \textcolor{black}{\textbf{$<$ 0.001}} \hspace{-0.1in} & \textcolor{black}{\textbf{$<$ 0.001}} \hspace{0.0in} &

\textcolor{black}{(9) w/ PT \textit{vs} (10) w/ PT} & \textcolor{black}{\textbf{0.004}} \hspace{-0.1in} & \textcolor{black}{\textbf{0.006}} \hspace{-0.1in} & \textcolor{black}{1} \hspace{-0.1in}
\\\noalign{\smallskip}\hline\noalign{\smallskip}

\textcolor{black}{(9) w/ PT \textit{vs} (10) w/o PT} & \textcolor{black}{1} \hspace{-0.1in} & \textcolor{black}{\textbf{0.017}} \hspace{-0.1in} & \textcolor{black}{0.256} \hspace{0.0in} &

\textcolor{black}{(9) w/o PT \textit{vs} (1) w/ PT} & \textcolor{black}{\textbf{$<$ 0.001}} \hspace{-0.1in} & \textcolor{black}{\textbf{$<$ 0.001}} \hspace{-0.1in} & \textcolor{black}{\textbf{$<$ 0.001}} \hspace{0.0in} &

\textcolor{black}{(9) w/o PT \textit{vs} (1) w/o PT} & \textcolor{black}{\textbf{$<$ 0.001}} \hspace{-0.1in} & \textcolor{black}{\textbf{$<$ 0.001}} \hspace{-0.1in} & \textcolor{black}{\textbf{$<$ 0.001}} \hspace{-0.1in}
\\\noalign{\smallskip}\hline\noalign{\smallskip}

\textcolor{black}{(9) w/o PT \textit{vs} (2) w/ PT} & \textcolor{black}{\textbf{$<$ 0.001}} \hspace{-0.1in} & \textcolor{black}{\textbf{$<$ 0.001}} \hspace{-0.1in} & \textcolor{black}{\textbf{$<$ 0.001}} \hspace{0.0in} &

\textcolor{black}{(9) w/o PT \textit{vs} (2) w/o PT} & \textcolor{black}{\textbf{$<$ 0.001}} \hspace{-0.1in} & \textcolor{black}{\textbf{$<$ 0.001}} \hspace{-0.1in} & \textcolor{black}{\textbf{$<$ 0.001}} \hspace{0.0in} &

\textcolor{black}{(9) w/o PT \textit{vs} (3) w/ PT} & \textcolor{black}{\textbf{$<$ 0.001}} \hspace{-0.1in} & \textcolor{black}{\textbf{$<$ 0.001}} \hspace{-0.1in} & \textcolor{black}{\textbf{$<$ 0.001}} \hspace{-0.1in}
\\\noalign{\smallskip}\hline\noalign{\smallskip}

\textcolor{black}{(9) w/o PT \textit{vs} (3) w/o PT} & \textcolor{black}{\textbf{$<$ 0.001}} \hspace{-0.1in} & \textcolor{black}{\textbf{$<$ 0.001}} \hspace{-0.1in} & \textcolor{black}{\textbf{$<$ 0.001}} \hspace{0.0in} &

\textcolor{black}{(9) w/o PT \textit{vs} (4) w/ PT} & \textcolor{black}{\textbf{$<$ 0.001}} \hspace{-0.1in} & \textcolor{black}{\textbf{$<$ 0.001}} \hspace{-0.1in} & \textcolor{black}{\textbf{$<$ 0.001}} \hspace{0.0in} &

\textcolor{black}{(9) w/o PT \textit{vs} (4) w/o PT} & \textcolor{black}{\textbf{$<$ 0.001}} \hspace{-0.1in} & \textcolor{black}{\textbf{$<$ 0.001}} \hspace{-0.1in} & \textcolor{black}{\textbf{$<$ 0.001}} \hspace{-0.1in}
\\\noalign{\smallskip}\hline\noalign{\smallskip}

\textcolor{black}{(9) w/o PT \textit{vs} (5) w/ PT} & \textcolor{black}{\textbf{$<$ 0.001}} \hspace{-0.1in} & \textcolor{black}{\textbf{$<$ 0.001}} \hspace{-0.1in} & \textcolor{black}{\textbf{$<$ 0.001}} \hspace{0.0in} &

\textcolor{black}{(9) w/o PT \textit{vs} (5) w/o PT} & \textcolor{black}{\textbf{0.022}} \hspace{-0.1in} & \textcolor{black}{1} \hspace{-0.1in} & \textcolor{black}{\textbf{$<$ 0.001}} \hspace{0.0in} &

\textcolor{black}{(9) w/o PT \textit{vs} (6) w/ PT} & \textcolor{black}{\textbf{$<$ 0.001}} \hspace{-0.1in} & \textcolor{black}{\textbf{$<$ 0.001}} \hspace{-0.1in} & \textcolor{black}{\textbf{$<$ 0.001}} \hspace{-0.1in}
\\\noalign{\smallskip}\hline\noalign{\smallskip}

\textcolor{black}{(9) w/o PT \textit{vs} (6) w/o PT} & \textcolor{black}{\textbf{$<$ 0.001}} \hspace{-0.1in} & \textcolor{black}{\textbf{$<$ 0.001}} \hspace{-0.1in} & \textcolor{black}{\textbf{$<$ 0.001}} \hspace{0.0in} &

\textcolor{black}{(9) w/o PT \textit{vs} (10) w/PT} & \textcolor{black}{\textbf{$<$ 0.001}} \hspace{-0.1in} & \textcolor{black}{\textbf{$<$ 0.001}} \hspace{-0.1in} & \textcolor{black}{\textbf{$<$ 0.001}} \hspace{0.0in} &

\textcolor{black}{(9) w/o PT \textit{vs} (10) w/o PT} & \textcolor{black}{\textbf{$<$ 0.001}} \hspace{-0.1in} & \textcolor{black}{\textbf{$<$ 0.001}} \hspace{-0.1in} & \textcolor{black}{\textbf{$<$ 0.001}} \hspace{-0.1in}
\\\noalign{\smallskip}\hline\noalign{\smallskip}

\textcolor{black}{(10) w/ PT \textit{vs} (1) w/ PT} & \textcolor{black}{\textbf{$<$ 0.001}} \hspace{-0.1in} & \textcolor{black}{\textbf{$<$ 0.001}} \hspace{-0.1in} & \textcolor{black}{\textbf{0.049}} \hspace{0.0in} &

\textcolor{black}{(10) w/ PT \textit{vs} (1) w/o PT} & \textcolor{black}{\textbf{$<$ 0.001}} \hspace{-0.1in} & \textcolor{black}{\textbf{$<$ 0.001}} \hspace{-0.1in} & \textcolor{black}{\textbf{$<$ 0.001}} \hspace{0.0in} &

\textcolor{black}{(10) w/ PT \textit{vs} (2) w/ PT} & \textcolor{black}{\textbf{0.039}} \hspace{-0.1in} & \textcolor{black}{0.515} \hspace{-0.1in} & \textcolor{black}{1} \hspace{-0.1in}
\\\noalign{\smallskip}\hline\noalign{\smallskip}

\textcolor{black}{(10) w/ PT \textit{vs} (2) w/o PT} & \textcolor{black}{\textbf{$<$ 0.001}} \hspace{-0.1in} & \textcolor{black}{\textbf{$<$ 0.001}} \hspace{-0.1in} & \textcolor{black}{\textbf{0.002}} \hspace{0.0in} &

\textcolor{black}{(10) w/ PT \textit{vs} (3) w/ PT} & \textcolor{black}{\textbf{0.017}} \hspace{-0.1in} & \textcolor{black}{\textbf{$<$ 0.001}} \hspace{-0.1in} & \textcolor{black}{1} \hspace{0.0in} &

\textcolor{black}{(10) w/ PT \textit{vs} (3) w/o PT} & \textcolor{black}{\textbf{$<$ 0.001}} \hspace{-0.1in} & \textcolor{black}{\textbf{$<$ 0.001}} \hspace{-0.1in} & \textcolor{black}{0.415} \hspace{-0.1in}
\\\noalign{\smallskip}\hline\noalign{\smallskip}

\textcolor{black}{(10) w/ PT \textit{vs} (4) w/ PT} & \textcolor{black}{\textbf{$<$ 0.001}} \hspace{-0.1in} & \textcolor{black}{\textbf{0.019}} \hspace{-0.1in} & \textcolor{black}{\textbf{0.028}} \hspace{0.0in} &

\textcolor{black}{(10) w/ PT \textit{vs} (4) w/o PT} & \textcolor{black}{\textbf{$<$ 0.001}} \hspace{-0.1in} & \textcolor{black}{\textbf{$<$ 0.001}} \hspace{-0.1in} & \textcolor{black}{\textbf{$<$ 0.001}} \hspace{0.0in} &

\textcolor{black}{(10) w/ PT \textit{vs} (5) w/ PT} & \textcolor{black}{\textbf{0.015}} \hspace{-0.1in} & \textcolor{black}{\textbf{$<$ 0.001}} \hspace{-0.1in} & \textcolor{black}{1} \hspace{-0.1in}
\\\noalign{\smallskip}\hline\noalign{\smallskip}

\textcolor{black}{(10) w/ PT \textit{vs} (5) w/o PT} & \textcolor{black}{\textbf{$<$ 0.001}} \hspace{-0.1in} & \textcolor{black}{\textbf{$<$ 0.001}} \hspace{-0.1in} & \textcolor{black}{\textbf{$<$ 0.001}} \hspace{0.0in} &

\textcolor{black}{(10) w/ PT \textit{vs} (6) w/ PT} & \textcolor{black}{1} \hspace{-0.1in} & \textcolor{black}{1} \hspace{-0.1in} & \textcolor{black}{1} \hspace{0.0in} &

\textcolor{black}{(10) w/ PT \textit{vs} (6) w/o PT} & \textcolor{black}{0.981} \hspace{-0.1in} & \textcolor{black}{1} \hspace{-0.1in} & \textcolor{black}{1} \hspace{-0.1in}
\\\noalign{\smallskip}\hline\noalign{\smallskip}

\textcolor{black}{(10) w/ PT \textit{vs} (10) w/o PT} & \textcolor{black}{0.054} \hspace{-0.1in} & \textcolor{black}{1} \hspace{-0.1in} & \textcolor{black}{\textbf{0.002}} \hspace{0.0in} &

\textcolor{black}{(10) w/o PT \textit{vs} (1) w/ PT} & \textcolor{black}{\textbf{0.039}} \hspace{-0.1in} & \textcolor{black}{\textbf{$<$ 0.001}} \hspace{-0.1in} & \textcolor{black}{1} \hspace{0.0in} &

\textcolor{black}{(10) w/o PT \textit{vs} (1) w/o PT} & \textcolor{black}{\textbf{$<$ 0.001}} \hspace{-0.1in} & \textcolor{black}{\textbf{$<$ 0.001}} \hspace{-0.1in} & \textcolor{black}{\textbf{$<$ 0.001}} \hspace{-0.1in}
\\\noalign{\smallskip}\hline\noalign{\smallskip}

\textcolor{black}{(10) w/o PT \textit{vs} (2) w/ PT} & \textcolor{black}{1} \hspace{-0.1in} & \textcolor{black}{0.727} \hspace{-0.1in} & \textcolor{black}{0.649} \hspace{0.0in} &

\textcolor{black}{(10) w/o PT \textit{vs} (2) w/o PT} & \textcolor{black}{\textbf{$<$ 0.001}} \hspace{-0.1in} & \textcolor{black}{\textbf{$<$ 0.001}} \hspace{-0.1in} & \textcolor{black}{1} \hspace{0.0in} &

\textcolor{black}{(10) w/o PT \textit{vs} (3) w/ PT} & \textcolor{black}{1} \hspace{-0.1in} & \textcolor{black}{\textbf{0.002}} \hspace{-0.1in} & \textcolor{black}{\textbf{$<$ 0.001}} \hspace{-0.1in}
\\\noalign{\smallskip}\hline\noalign{\smallskip}

\textcolor{black}{(10) w/o PT \textit{vs} (3) w/o PT} & \textcolor{black}{\textbf{0.039}} \hspace{-0.1in} & \textcolor{black}{\textbf{$<$ 0.001}} \hspace{-0.1in} & \textcolor{black}{1} \hspace{0.0in} &

\textcolor{black}{(10) w/o PT \textit{vs} (4) w/ PT} & \textcolor{black}{1} \hspace{-0.1in} & \textcolor{black}{\textbf{0.019}} \hspace{-0.1in} & \textcolor{black}{1} \hspace{0.0in} &

\textcolor{black}{(10) w/o PT \textit{vs} (4) w/o PT} & \textcolor{black}{\textbf{$<$ 0.001}} \hspace{-0.1in} & \textcolor{black}{\textbf{$<$ 0.001}} \hspace{-0.1in} & \textcolor{black}{\textbf{$<$ 0.001}} \hspace{-0.1in}
\\\noalign{\smallskip}\hline\noalign{\smallskip}

\textcolor{black}{(10) w/o PT \textit{vs} (5) w/ PT} & \textcolor{black}{1} \hspace{-0.1in} & \textcolor{black}{\textbf{0.002}} \hspace{-0.1in} & \textcolor{black}{\textbf{$<$ 0.001}} \hspace{0.0in} &

\textcolor{black}{(10) w/o PT \textit{vs} (5) w/o PT} & \textcolor{black}{\textbf{$<$ 0.001}} \hspace{-0.1in} & \textcolor{black}{\textbf{$<$ 0.001}} \hspace{-0.1in} & \textcolor{black}{\textbf{$<$ 0.001}} \hspace{0.0in} &

\textcolor{black}{(10) w/o PT \textit{vs} (6) w/ PT} & \textcolor{black}{0.308} \hspace{-0.1in} & \textcolor{black}{1} \hspace{-0.1in} & \textcolor{black}{\textbf{$<$ 0.001}} \hspace{-0.1in}
\\\noalign{\smallskip}\hline\noalign{\smallskip}

\textcolor{black}{(10) w/o PT \textit{vs} (6) w/o PT} & \textcolor{black}{1} \hspace{-0.1in} & \textcolor{black}{1} \hspace{-0.1in} & \textcolor{black}{\textbf{0.035}} \hspace{0.0in} &
\\\noalign{\smallskip}\Hline\noalign{\smallskip}

\label{tab6}
\end{tabular}
}
\end{small}
\vspace{-0.1in}
\end{table*}

\subsection{Clinical Validation Using Visual Turing Test}
To \textcolor{black}{quantify} the (\textit{i}) realism of \textcolor{black}{$224 \times 224$} synthetic images \textcolor{black}{by PGGANs, \textcolor{black}{MUNIT}, and SimGAN against real images respectively (i.e., 3 setups)} and (\textit{ii}) clearness of their tumor/non-tumor features, we supply, in random order, to an expert physician a random selection of:
\begin{itemize}
\item $50$ real tumor images;
\item $50$ real non-tumor images;
\item $50$ synthetic tumor images;
\item $50$ synthetic non-tumor images.
\end{itemize}

Then, the physician has to classify them as both (\textit{i}) real/synthetic and (\textit{ii}) tumor/non-tumor, without previously knowing which is real/synthetic and tumor/non-tumor.
The so-called Visual Turing Test~\cite{Salimans} can probe the human ability to identify attributes and relationships in images, also for visually evaluating GAN-generated images~\cite{Shrivastava}; this also applies to medical images for clinical decision-making tasks~\cite{Chuquicusma,Han2}, wherein physicians' expertise is critical.

\subsection{Visualization Using t-SNE}
To \textcolor{black}{visualize} distributions of geometrically-transformed and each GAN-based \textcolor{black}{$224 \times 224$} images by PGGANs, \textcolor{black}{MUNIT}, and SimGAN against real images \textcolor{black}{respectively} (i.e., 4 setups), we adopt t-SNE~\cite{van der Maaten} on a random selection of:
\begin{itemize}
\item $300$ real tumor images;
\item $300$ real non-tumor images;
\item $300$ geometrically-transformed or each GAN-based tumor images;
\item $300$ geometrically-transformed or each GAN-based non-tumor images.
\end{itemize}

We select only $300$ images per each category for better visualization. The t-SNE method reduces the dimensionality to represent high-dimensional data into a lower-dimensional (2D/3D) space; it non-linearly balances between the input data's local and global aspects using perplexity.

\vspace{0.1in}

\noindent \textbf{T-SNE Implementation Details}
The t-SNE uses a perplexity of $100$ for $1,000$ iterations to visually represent a 2D space.
\textcolor{black}{We input the images after normalizing pixel values to $[0, 1]$. For point locations of the real images, we compress all the images simultaneously and plot each setup (i.e., the geometrically-transformed or each GAN-based images against the real ones) separately; we maintain their locations by projecting all the data onto the same subspace.}

\begin{figure}[t]
  \centering
  \centerline{\includegraphics[width=\columnwidth]{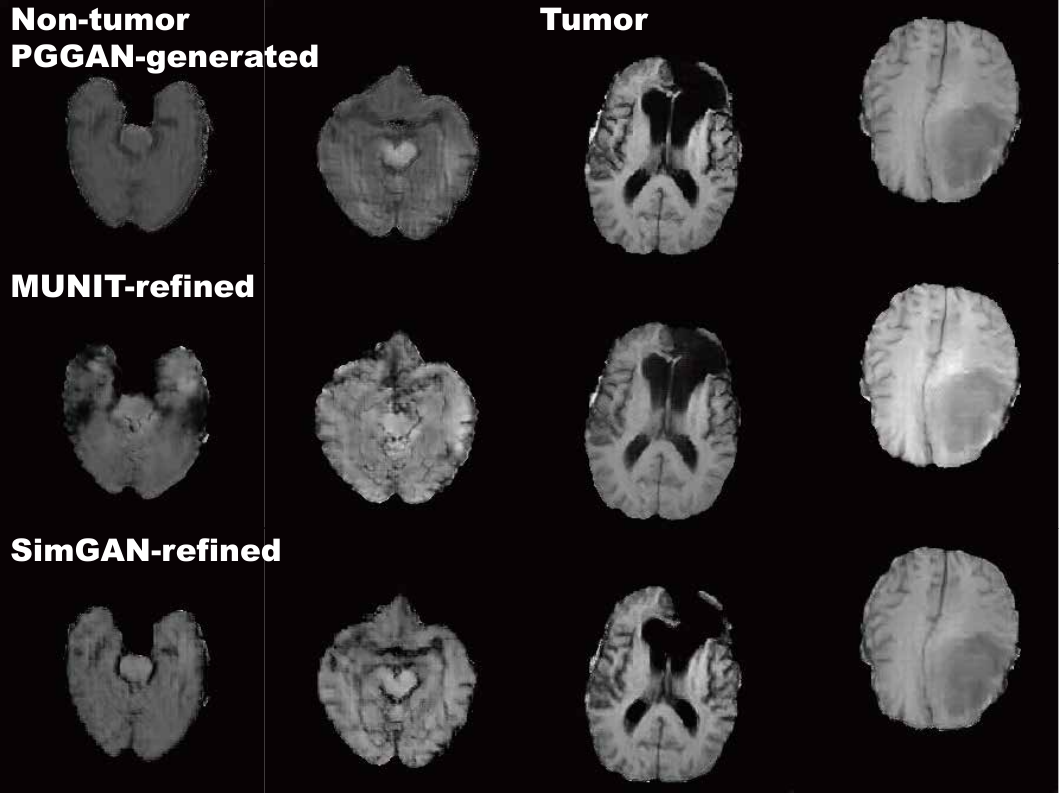}}
\caption{Example PGGAN-generated MR images and their refined versions by \textcolor{black}{MUNIT}/SimGAN.}
\label{fig6}
\end{figure}

\section{Results}
\label{sec:results}
This section shows how PGGANs generates synthetic brain MR images and how \textcolor{black}{MUNIT} and SimGAN refine them. The results include instances of synthetic images, their quantitative evaluation by an expert physician, their t-SNE visualization, and their influence on tumor detection.

\subsection{MR Images Generated by PGGANs}
Fig.~\ref{fig5} illustrates examples of synthetic MR images by PGGANs. We visually confirm that, for about $75\%$ of cases, it successfully captures the T1c-specific texture and tumor appearance, while maintaining the realism of the original brain MR images; but, for the rest $25\%$, the generated images lack clear tumor/non-tumor features or contain unrealistic features (i.e., hyper-intensity, gray contours, and odd artifacts).

\subsection{MR Images Refined by MUNIT/SimGAN}
\textcolor{black}{MUNIT} and SimGAN differently refine PGGAN-generated images---they render the texture and contours while maintaining the overall shape (Fig.~\ref{fig6}). Non-tumor images change more remarkably than tumor images for both \textcolor{black}{MUNIT} and SimGAN; it probably derives from unsupervised image translation's loss for consistency to avoid image collapse, resulting in conservative change for more complicated images.

\begin{table*}[!t]
\caption{Visual Turing Test results by an expert physician for classifying Real ($\mathsf{R}$) \textit{vs} Synthetic ($\mathsf{S}$) images and Tumor ($\mathsf{T}$) \textit{vs} Non-tumor ($\mathsf{N}$) images. \textcolor{black}{Accuracy denotes the physician's successful classification ratio between the real/synthetic images and between the tumor/non-tumor images, respectively. It should be noted that proximity to 50\% of accuracy indicates superior performance (chance = 50\%).}}
\label{tab7}
\centering
\begin{small}
\begin{tabular}{p{1.4em}ccccc}
\Hline\noalign{\smallskip}
\parbox[t]{2mm}{\multirow{4}{*}{\rotatebox[origin=c]{270}{\textbf{\shortstack{\textcolor{black}{PGGAN}}}}}} & \multicolumn{1}{c}{\textcolor{black}{Accuracy (Real \textit{vs} Synthetic)}}  \ \ \ &  $\mathsf{R}$ as $\mathsf{R}$ \ \ \ &  $\mathsf{R}$ as $\mathsf{S}$ \ \ \ &  $\mathsf{S}$ as $\mathsf{R}$ \ \ \ &  $\mathsf{S}$ as $\mathsf{S}$\\ 
& $79.5\%$ \ \ \ & $73\textcolor{black}{\%}$ \ \ \ & $27\textcolor{black}{\%}$ \ \ \ & $14\textcolor{black}{\%}$ \ \ \ & $86\textcolor{black}{\%}$ \\
& \multicolumn{1}{c}{\textcolor{black}{Accuracy (Tumor \textit{vs} Non-tumor)}} \ \ \ &  $\mathsf{T}$ as $\mathsf{T}$ \ \ \ &  $\mathsf{T}$ as $\mathsf{N}$ \ \ \ &  $\mathsf{N}$ as $\mathsf{T}$ \ \ \ &  $\mathsf{N}$ as $\mathsf{N}$\\
& $87.5\%$ \ \ \ & $77\textcolor{black}{\%}$ \ \ \ & $23\textcolor{black}{\%}$ ($\mathsf{R}: 
11$, $\mathsf{S}: 12$) \ \ \ & $2\textcolor{black}{\%}$ ($\mathsf{S}: 2$) \ \ \ & $98\textcolor{black}{\%}$\\
\noalign{\smallskip}\hline\noalign{\smallskip}
\parbox[t]{2mm}{\multirow{4}{*}{\rotatebox[origin=c]{270}{\textbf{\shortstack{\textcolor{black}{MUNIT}}}}}} & \multicolumn{1}{c}{\textcolor{black}{Accuracy (Real \textit{vs} Synthetic)}}  \ \ \ &  \textcolor{black}{$\mathsf{R}$ as $\mathsf{R}$} \ \ \ &  \textcolor{black}{$\mathsf{R}$ as $\mathsf{S}$} \ \ \ &  \textcolor{black}{$\mathsf{S}$ as $\mathsf{R}$} \ \ \ &  \textcolor{black}{$\mathsf{S}$ as $\mathsf{S}$}\\ 

& \textcolor{black}{$77.0\%$} \ \ \ & \textcolor{black}{$58\%$} \ \ \ & \textcolor{black}{$42\%$} \ \ \ & \textcolor{black}{$4\%$} \ \ \ & \textcolor{black}{$96\%$} \\
& \multicolumn{1}{c}{\textcolor{black}{Accuracy (Tumor \textit{vs} Non-tumor)}} \ \ \ &  \textcolor{black}{$\mathsf{T}$ as $\mathsf{T}$} \ \ \ &  \textcolor{black}{$\mathsf{T}$ as $\mathsf{N}$} \ \ \ &  \textcolor{black}{$\mathsf{N}$ as $\mathsf{T}$} \ \ \ &  \textcolor{black}{$\mathsf{N}$ as $\mathsf{N}$}\\
& \textcolor{black}{$92.5\%$} \ \ \ & \textcolor{black}{$88\%$} \ \ \ & \textcolor{black}{$12\%$ ($\mathsf{R}: 
6$, $\mathsf{S}: 6$)} \ \ \ & \textcolor{black}{$3\%$ ($\mathsf{R}: 1$, $\mathsf{S}: 2$)} \ \ \ & \textcolor{black}{$97\%$}\\
\noalign{\smallskip}\hline\noalign{\smallskip}
\parbox[t]{2mm}{\multirow{4}{*}{\rotatebox[origin=c]{270}{\textbf{\shortstack{\textcolor{black}{SimGAN}}}}}} & \multicolumn{1}{c}{\textcolor{black}{Accuracy (Real \textit{vs} Synthetic})}  \ \ \ &  \textcolor{black}{$\mathsf{R}$ as $\mathsf{R}$} \ \ \ &  \textcolor{black}{$\mathsf{R}$ as $\mathsf{S}$} \ \ \ &  \textcolor{black}{$\mathsf{S}$ as $\mathsf{R}$} \ \ \ &  \textcolor{black}{$\mathsf{S}$ as $\mathsf{S}$}\\ 
& \textcolor{black}{$76.0\%$} \ \ \ & \textcolor{black}{$53\%$} \ \ \ & \textcolor{black}{$47\%$} \ \ \ & \textcolor{black}{$1\%$} \ \ \ & \textcolor{black}{$99\%$} \\
& \multicolumn{1}{c}{\textcolor{black}{Accuracy (Tumor \textit{vs} Non-tumor)}} \ \ \ &  \textcolor{black}{$\mathsf{T}$ as $\mathsf{T}$} \ \ \ &  \textcolor{black}{$\mathsf{T}$ as $\mathsf{N}$} \ \ \ &  \textcolor{black}{$\mathsf{N}$ as $\mathsf{T}$} \ \ \ &  \textcolor{black}{$\mathsf{N}$ as $\mathsf{N}$}\\
& \textcolor{black}{$94.0\%$} \ \ \ & \textcolor{black}{$91\%$} \ \ \ & \textcolor{black}{$9\%$ ($\mathsf{R}: 
2$, $\mathsf{S}: 7$)} \ \ \ & \textcolor{black}{$3\%$ ($\mathsf{R}: 3$)} \ \ \ & \textcolor{black}{$97\%$}\\
\noalign{\smallskip}\Hline\noalign{\smallskip}
\end{tabular}
\end{small}
\end{table*}

\begin{figure*}[t]
  \centering
  \centerline{\includegraphics[width=2\columnwidth]{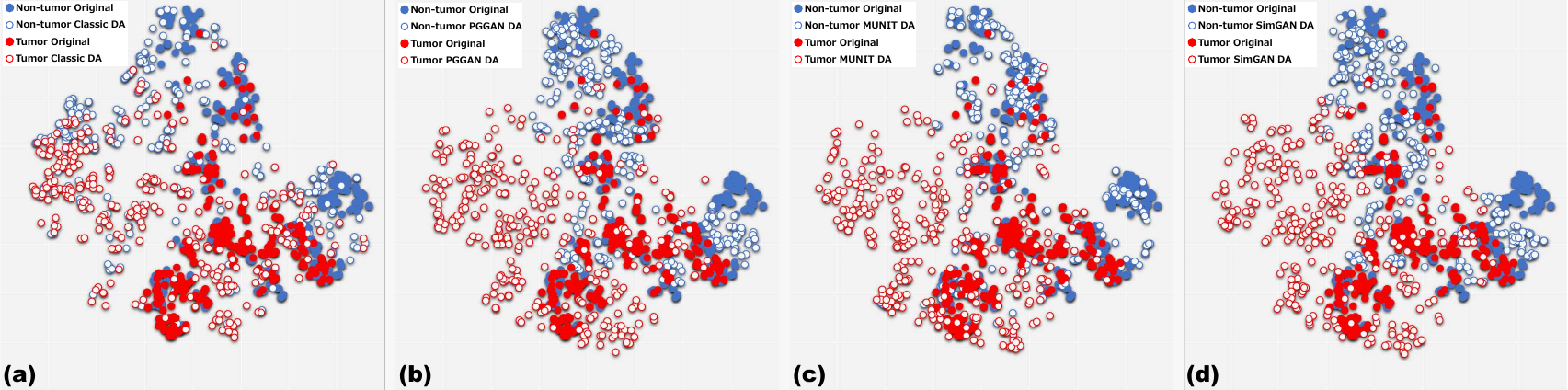}}
\caption{T-SNE plots with 300 tumor/non-tumor MR images per each category: Real images \textit{vs} (a) Geometrically-transformed images; (b) PGGAN-generated images; (c) \textcolor{black}{MUNIT}-refined images; (d) SimGAN-refined images.}
\label{fig7}
\end{figure*}

\subsection{Tumor Detection Results}

Table~\ref{tab5} shows the brain tumor classification results with/without DA \textcolor{black}{while Table~\ref{tab6} indicates their pairwise comparison ($p$-values between our two-step GAN-based DA setups and the other DA setups) using McNemar's test.} ImageNet pre-training generally outperforms training from scratch despite different image domains (i.e., natural images to medical images). As expected, classic DA remarkably improves classification, while no clear difference exists between the $200,000$/$400,000$ classic DA under sufficient geometrically-transformed training images. When pre-trained, each GAN-based DA (i.e., PGGANs/\textcolor{black}{MUNIT}/SimGAN) alone helps classification due to the robustness from GAN-generated images; but, without pre-training, it harms classification due to the biased initialization from the GAN-overwhelming data distribution. Similarly, without pre-training, PGGAN-based DA without clustering/discarding causes poor classification due to the synthetic images with severe artifacts, unlike the PGGAN-based DA's comparable results with/without the discarding step when pre-trained.

When combined with the classic DA, each GAN-based DA remarkably outperforms the GAN-based DA or classic DA alone \textcolor{black}{in terms of sensitivity since they are mutually-complementary: the former learns the non-linear manifold of the real images to generate novel local tumor features (since we train tumor/non-tumor images separately) strongly associated with sensitivity; the latter learns the geometrically-transformed manifold of the real images to cover global features and provide the robustness on training for most cases. We confirm that test samples, originally-misclassified but correctly classified after DA, are obviously different for the GAN-based DA and classic DA;} here, both image-to-image GAN-based DA, especially \textcolor{black}{MUNIT}, produce remarkably higher sensitivity than the PGGAN-based DA after refinement. Specificity is higher than sensitivity for every DA setup with pre-training, probably due to the training data imbalance; but interestingly, without pre-training, sensitivity is higher than specificity for both image-to-image GAN-based DA \textcolor{black}{since our tumor detection-oriented two-step GAN-based DA can fill the real tumor image distribution uncovered by the original dataset under no ImageNet initialization. Accordingly,} when combined with the classic DA, the \textcolor{black}{MUNIT}-based DA \textcolor{black}{based on both GANs/VAEs} achieves the highest sensitivity \textcolor{black}{$97.48\%$ against the best performing classic DA's $93.67\%$}, allowing to significantly alleviate the risk of overlooking the tumor diagnosis; \textcolor{black}{in terms of sensitivity, it outperforms all the other DA setups, including two-step DA setups, with statistical significance.}


\subsection{Visual Turing Test Results}
Table~\ref{tab7} indicates the confusion matrix for the Visual Turing Test. The expert physician classifies a few PGGAN-generated images as real\textcolor{black}{, thanks to their realism,} despite high resolution (i.e., \textcolor{black}{$224 \times 224$} pixels); \textcolor{black}{meanwhile, the expert classifies less GAN-refined images as real due to slight artifacts induced during refinement.} The synthetic images successfully capture tumor/non-tumor features; unlike the non-tumor images, the expert recognizes a considerable number of the mild/modest tumor images as non-tumor for both real/synthetic cases. It derives from clinical tumor diagnosis relying on a full 3D volume, instead of a single 2D slice.

\subsection{T-SNE Results}
As Fig.~\ref{fig7} represents, the real tumor/non-tumor image distributions largely overlap while the non-tumor images distribute wider. The geometrically-transformed tumor/non-tumor image distributions also often overlap, and both images distribute wider than the real ones. All GAN-based synthetic images by PGGANs/\textcolor{black}{MUNIT}/SimGAN distribute widely, while their tumor/non-tumor images overlap much less than the geometrically-transformed ones \textcolor{black}{(i.e., a high discrimination ability associated with sensitivity improvement)}; the \textcolor{black}{MUNIT}-refined images show \textcolor{black}{better tumor/non-tumor discrimination and} a more similar distribution to the real ones than the PGGAN/SimGAN-based images. \textcolor{black}{This trend derives from} the \textcolor{black}{MUNIT}'s loss function adopting both GANs/VAEs \textcolor{black}{that further fits the PGGAN-generated images into the real image distribution by refining their texture/shape; contrarily, this refinement could also induce slight human-recognizable but DA-irrelevant artifacts}. Overall, the GAN-based images, especially the \textcolor{black}{MUNIT}-refined images, fill the distribution uncovered by the real or geometrically-transformed ones with less tumor/non-tumor overlap; \textcolor{black}{this demonstrates the superiority of combining classic DA and GAN-based DA}.

\section{Conclusion}
\label{sec:conclusions}
Visual Turing Test and t-SNE results show that PGGANs, multi-stage noise-to-image GAN, can generate realistic/diverse $256 \times 256$ brain MR images with/without tumors separately. \textcolor{black}{Unlike classic DA that geometrically covers global features and provides the robustness on training for most cases, the GAN-generated images can non-linearly cover local tumor features with much less tumor/non-tumor overlap; thus, combining them can significantly boost tumor detection sensitivity}---especially after refining them with \textcolor{black}{MUNIT} or SimGAN, image-to-image GANs; thanks to an ensemble \textcolor{black}{generation process} from those GANs' different algorithms, the \textcolor{black}{texture/shape-}refined images can replace missing data points of the training set with less tumor/non-tumor overlap, and thus handle the data imbalance \textcolor{black}{by regularizing the model (i.e., improved generalization). Notably, MUNIT remarkably} outperforms SimGAN \textcolor{black}{in terms of sensitivity}, probably due to the effect of combining both GANs/VAEs.

Regarding better medical GAN-based DA, ImageNet pre-training generally improves classification despite different textures of natural/medical images; but, without pre-training, the GAN-refined images may help achieve better sensitivity, allowing to alleviate the risk of overlooking the tumor diagnosis\textcolor{black}{---this attributes to our tumor detection-oriented two-step GAN-based DA's high discrimination ability to fill the real tumor image distribution under no ImageNet initialization.} GAN-generated images typically include odd artifacts; however, only without pre-training, discarding them boosts DA performance.

Overall, by minimizing the number of annotated images required for medical imaging tasks, the two-step GAN-based DA can shed light not only on classification, but also on object detection~\cite{Han3} and segmentation~\cite{Shin}. Moreover, other potential medical applications exist: (\textit{i}) A data anonymization tool to share patients' data outside their institution for training without losing detection performance~\cite{Shin}; (\textit{ii}) A physician training tool to show random pathological images for medical students/radiology trainees despite infrastructural/legal constraints~\cite{Finlayson}. As future work, we plan to define a new \textcolor{black}{end-to-end} GAN loss function that explicitly \textcolor{black}{optimizes} the classification results, instead of \textcolor{black}{optimizing} visual realism \textcolor{black}{while maintaining diversity by combining the state-of-the-art noise-to-image and image-to-image GANs; towards this, we might extend a preliminary work on a three-player GAN for classification~\cite{Vandenhende} to generate only hard-to-classify samples to improve classification; we could also (\textit{i}) explicitly model deformation fields/intensity transformations and (\textit{ii}) leverage unlabelled data during the generative process~\cite{Chaitanya} to effectively fill the real image distribution.}

\section*{Acknowledgment}
This research was partially supported by Qdai-jump Research Program, JSPS KAKENHI Grant Number JP17K12752, and AMED Grant Number JP18lk1010028.

\newpage

\begin{IEEEbiography}[{\includegraphics[width=1in,height=1.25in,clip,keepaspectratio]{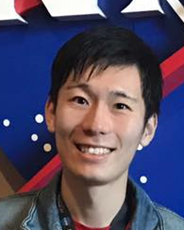}}]{CHANGHEE HAN} received the Bachelor’s and Master’s Degrees in Computer Science from The University of Tokyo in 2015 and 2017, respectively. Since 2017, he is a Ph.D. student at Graduate School of Information Science and Technology, The University of Tokyo. He is currently a Visiting Scholar at National Center for Global Health and Medicine since 2018. He was invited as a Visiting Scholar at the Technical University of Munich in 2016, University of Milano-Bicocca in 2017 and 2018, and University of Cambridge in 2019. His research interests include Machine Learning, especially Deep Learning for Medical Imaging and Bioinformatics.
\end{IEEEbiography}

\begin{IEEEbiography}[{\includegraphics[width=1in,height=1.25in,clip,keepaspectratio]{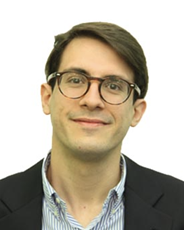}}]{LEONARDO RUNDO} received the Bachelor’s and Master’s Degrees in Computer Science Engineering from the University of Palermo in 2010 and 2013, respectively. In 2013, he was a Research Fellow at the Institute of Molecular Bioimaging and Physiology, National Research Council of Italy (IBFM-CNR). He obtained his Ph.D. in Computer Science at the University of Milano-Bicocca in 2019. Since November 2018, he is a Research Associate at the Department of Radiology, University of Cambridge, collaborating with Cancer Research UK. His main scientific interests include Biomedical Image Analysis, Machine Learning, Computational Intelligence, and High-Performance Computing.
\end{IEEEbiography}

\begin{IEEEbiography}[{\includegraphics[width=1in,height=1.25in,clip,keepaspectratio]{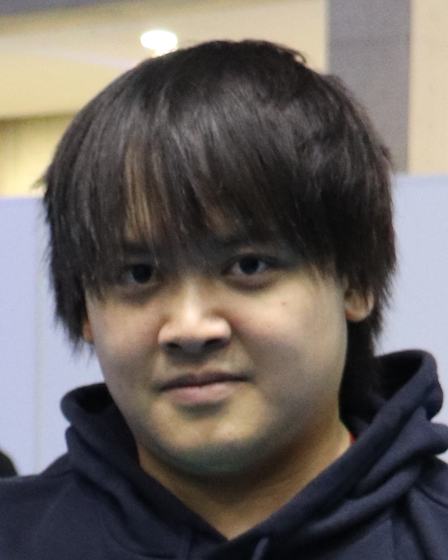}}]{RYOSUKE ARAKI} received the Bachelor’s and Master’s Degrees in Engineering from Chubu University, in 2017 and 2019, respectively. Since 2019, he is a Ph.D. student at Graduate School of Engineering, Chubu University. His research interests include Computer Vision and Robot Vision, especially Deep Learning for Intelligent Robotics.
\end{IEEEbiography}

\begin{IEEEbiography}[{\includegraphics[width=1in,height=1.25in,clip,keepaspectratio]{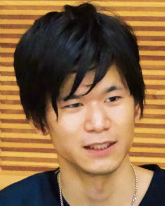}}]{YUDAI NAGANO} received the Master's Degree in Computer Science from the University of Tokyo in 2019. Since 2019, he is a Ph.D. student at Graduate school of Information Science and Technology, the University of Tokyo. His main research interests include Generative Adversarial Networks for super-resolution, image-to-image translation, and segmentation.
\end{IEEEbiography}

\begin{IEEEbiography}[{\includegraphics[width=1in,height=1.25in,clip,keepaspectratio]{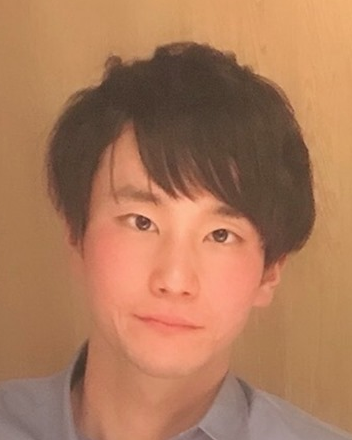}}]{YUJIRO FURUKAWA} received the M.D. from Akita University in 2015. Since 2019, he is a Psychiatrist at the Jikei University Hospital. His research interests include Medical Imaging of Dementia and Depression.
\end{IEEEbiography}

\begin{IEEEbiography}[{\includegraphics[width=1in,height=1.25in,clip,keepaspectratio]{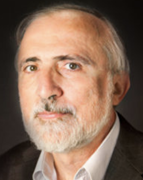}}]{GIANCARLO MAURI} is a Full Professor of Computer Science at University of Milano-Bicocca. His research interests include: natural computing and unconventional computing models, bioinformatics, stochastic modeling, and simulation of biological systems and processes using High-Performance Computing approaches, biomedical data analysis. On these subjects, he published about 400 scientific papers. He is a member of the steering committees of the International Conference on Membrane Computing, of the International Conference on Unconventional Computing and Natural Computing and, of the International workshop on Cellular Automata.
\end{IEEEbiography}

\begin{IEEEbiography}[{\includegraphics[width=1in,height=1.25in,clip,keepaspectratio]{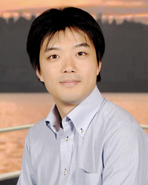}}]{HIDEKI NAKAYAMA} received the Master's and Ph.D. Degrees in Information Science from the University of Tokyo, Japan in 2008 and 2011, respectively. From 2012 to 2018, he was an Assistant Professor at the Graduate School of Information Science and Technology, the University of Tokyo, Japan. Since April 2018, he has been an Associate Professor at the same department. He is also an affiliated faculty of the International Research Center for Neurointelligence (IRCN), and a Visiting Researcher at the National Institute of Advanced Industrial Science and Technology (AIST). His research interests include generic image recognition, Natural Language Processing, and Deep Learning.
\end{IEEEbiography}

\begin{IEEEbiography}[{\includegraphics[width=1in,height=1.25in,clip,keepaspectratio]{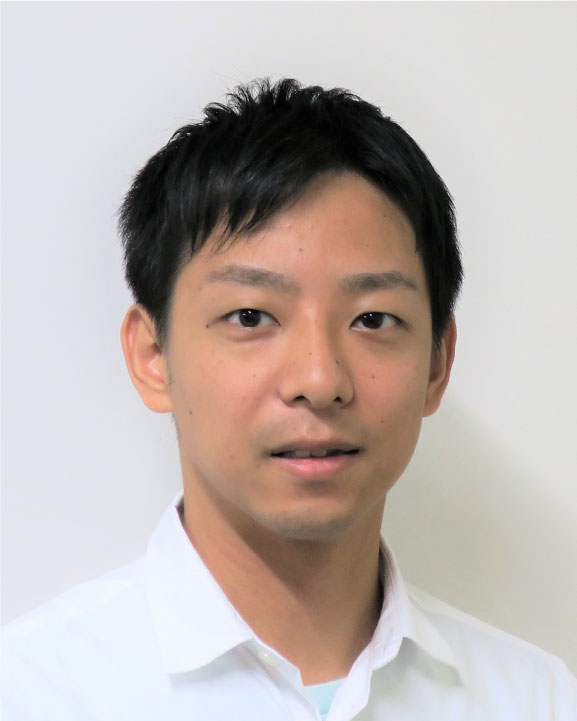}}]{HIDEAKI HAYASHI} received the Bachelor’s, Master’s, and Ph.D. degrees in Engineering from Hiroshima University in 2012, 2014, and 2016, respectively. He was a Research Fellow of the Japan Society for the Promotion of Science from 2015 to 2017. He is currently an assistant professor in the Department of Advanced Information Technology, Kyushu University. His research interests include biosignal analysis, neural networks, and machine learning.
\end{IEEEbiography}

\EOD

\end{document}